\documentclass{jfm}
\usepackage{graphicx}
\usepackage[export]{adjustbox}
\usepackage[skip=3pt]{caption, subcaption}
\usepackage{xcolor}
\usepackage{hyperref}

\title{Viscous jetting and Mach stem bifurcation in shock reflections: experiments and simulations}
\shorttitle{Jetting and Mach stem bifurcation in shock reflections}
\shortauthor{S. S.-M. Lau-Chapdelaine, Q. Xiao and M. I. Radulescu}

\author{S. S.-M. Lau-Chapdelaine\aff{1}
  \corresp{\email{Shem.Lau-Chapdelaine@rmc.ca}},
  Q. Xiao\aff{2}
 \and M. I. Radulescu\aff{2}}

\affiliation{
	\aff{1}Department of Chemistry and Chemical Engineering, Royal Military College, 11 Crerar Cres., Kingston, Ontario,  K7K 7B4, Canada
	\aff{2}Department of Mechanical Engineering, University of Ottawa, 161 Louis Pasteur Pvt., Ottawa, Ontario, K1N 6N5, Canada}

\newcommand{\micro}{$\mu$}
\newcommand{\EXP}[1]{{\times}10^{#1}}

\begin{document}

\maketitle

\begin{abstract}
Shock reflection experiments are performed to study the large-scale convective mixing created by the forward jetting phenomenon. Experiments are performed at a wedge angle of $\theta_{\mathrm{w}} = 30^{\circ}$ in nitrogen, propane-oxygen, and hexane with incident shock Mach numbers up to $M = 4$. Experiments are complimented by shock-resolved viscous simulations of triple point reflection in hexane for $M = 2.5$ to $6$. Inviscid simulations are performed over a wider range of parameters. Reynolds numbers up to $Re \lesssim 10^3$ are covered by simulations and Reynolds numbers of $Re \sim 10^5$ are covered by experiments.
The study shows that as the isentropic exponent is lowered, and as the Mach number and Reynolds number are increased, the forward jet approaches the Mach stem, forms a vortex, deforms the shock front and in some cases bifurcates the Mach stem. Experiments show Kelvin-Helmholtz instabilities in the vortex cause large-scale convective mixing behind the Mach stem at low isentropic exponents ($\gamma \le 1.15$).
The limits of Mach stem bifurcation (triple Mach-White reflection) in inviscid simulations are plotted in the phase space of $M$- $\theta_{\mathrm{w}}$-$\gamma$. A maximum isentropic exponent of $\gamma \approx 1.3$ is found beyond which bifurcation does not occur (at $\theta_{\mathrm{w}} = 30^{\circ}$). This closely matches the boundary between irregular/regular detonation cellular structures.

\end{abstract}


\section{Introduction}\label{sec:introduction}

\begin{figure}
	\centering
	\includegraphics[scale=1]{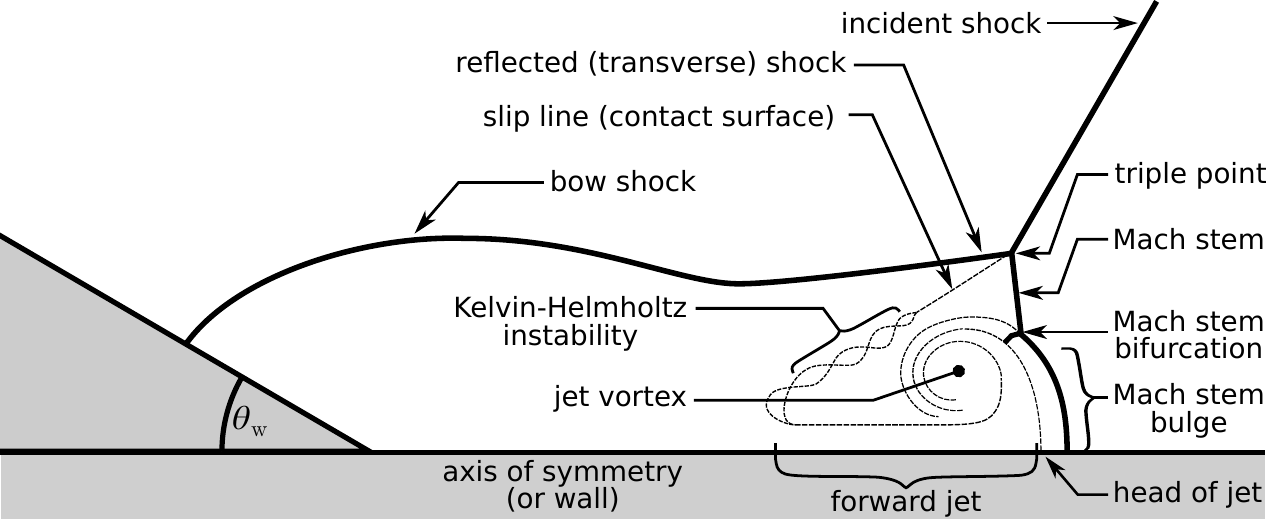}
	\caption{\label{fig:Mach reflection}Illustration of a Mach reflection with Mach stem bifurcation}
\end{figure}

One major category of shock reflection is the Mach reflection, illustrated in figure \ref{fig:Mach reflection}. It is composed of three shocks: the incident, reflected, and Mach shocks, and a slip line joined at the triple point. The slip line separates gas shocked by the incident and reflected shocks from gas shocked by the Mach shock (Mach stem). Under certain conditions (\cite{hornung_regular_1986,henderson_wall-jetting_2003}), the slip line curls towards the Mach stem and forms a forward jet. As the isentropic exponent is lowered or the Mach number is increased (\cite{mach_bifurcating_2011}), the head of the forward jet approaches the Mach stem, eventually causing it to bulge, kink, or even bifurcate, meaning a new triple point is formed on the Mach stem. The jet rolls up into a large vortex behind the Mach stem. 

The configuration of shock reflection depends on three parameters: the Mach number $M$, angle of reflection (\textit{i.e.} wedge angle $\theta_{\mathrm{w}}$), and the isentropic exponent (heat capacity ratio) $\gamma$. Classifications of shock reflection types are presented in the works of \cite{ben-dor_shock_2007,vasilev_analytical_2008,semenov_classification_2012} and thorough reviews on shock reflections have been written by \cite{hornung_regular_1986} and \cite{ben-dor_shock_2007}.

The parameters that lead to forward jet and vortex formation are typical of hydrocarbon detonations (\textit{i.e.} low $\gamma$, high $M$). These phenomena are important to understanding the multi-dimensional instability and propagation limits of detonations. 
For instance, the forward jet (\cite{sorin_detonation_2009}) and vortex are believed to accelerate reaction rates through large-scale mixing (\cite{maley_influence_2015}), while Mach stem bifurcation caused by the jet is postulated (\cite{radulescu_origin_2009}) to be a source of new detonation cells.

In this study, inert shock reflection simulations and experiments are used to investigate the cause of large-scale mixing in shock reflections as they pertain to detonations. Particular attention will be placed on the role of the forward jet and the jet-shock interaction. The simulations and experiments are designed to overcome some of the challenges posed by traditional wedge reflection configurations.

There is conflicting evidence on the prominence of the jet and vortex, or the conditions under which they affect the shape of the shock front.
Comparisons of experiments and inviscid simulations (\cite{glaz_detailed_1985,glaz_numerical_1985}) have found that the Mach stem bulges more in simulations than experiments.
Bulging can become severe enough to cause Mach stem bifurcation in inviscid simulations (\textit{e.g.} \cite{mach_bifurcating_2011}) but has only been observed in a few experiments (\cite{semenov_classification_2009b,maley_influence_2015,mach_mach_2011}). 

The fundamental issue with the Euler equations, often used to model shock reflections and detonations, is that they do not converge in multi-dimensional cases featuring slip lines (\cite{samtaney_initial-value_1996}). The shock reflection and forward jet behave differently depending on the numerical scheme and grid layout (\cite{quirk_contribution_1992,lau-chapdelaine_non-uniqueness_2013}) as they differ in how artificial diffusion is introduced. The shortcomings of the Euler equations are overcome in this study by use of the Navier-Stokes equations.

Implementation of the reflecting surface also poses problems. 
In simulations, surfaces oblique to the grid alter the reflection configuration depending on how they are implemented (\cite{ben-dor_influence_1987,lau-chapdelaine_non-uniqueness_2013}). This will be avoided by using triple point reflections (\cite{lau-chapdelaine_viscous_2016}) from grid-aligned surfaces.

Boundary layers on the reflecting surface change the effective wedge angle (\cite{hornung_effect_1985,ben-dor_shock_2007}) and cause the forward jet to oscillate (\cite{vasilev_wall-jetting_2004,shi_numerical_2017}). 
While most simulations use free-slip boundaries, the few experiments (\cite{smith_mutual_1959,henderson_experiments_1975,higashino_experiments_1991,barbosa_experimental_2002}) performed with free-slip surfaces were focused on the transition from regular to irregular reflection.
Shock reflections in detonations happen on a symmetry boundary, where boundary layers are absent, and there is a need for experiments to explore jetting under these conditions.

In this study, well posed experiments and simulations are used to explore the conditions that lead to a strong forward jet, vortex, and jet-shock interaction. This is done through experiments of shock reflection from a plane of symmetry, shock-resolved Navier-Stokes simulations, and inviscid simulations. Viscous simulations are performed at Reynolds numbers up to $Re \sim 10^3$ and experiments up to $\sim 10^5$. For comparison, detonations in stoichiometric hydrogen-, methane-, and propane-oxygen at atmospheric conditions have been estimated to reach Reynolds numbers of $\sim 10^5$ to $10^6$ on cellular length scales (\cite{lau-chapdelaine_viscous_2016}). The effects of Reynolds number, Mach number, and isentropic exponent on the forward jet, flow field behind the Mach stem, and the shock front are scrutinized. Conditions where jet-shock interaction become important are determined and their boundaries are reported.

The experiments are addressed in section \ref{sec:experiments} followed by numerical simulations in section \ref{sec:numerical prediction}. The discussion is found in section \ref{sec:discussion} and the conclusion in section \ref{sec:conclusion}.

\section{Experiments}\label{sec:experiments}

\subsection{Experimental technique}

\begin{figure}
	\centering
	\begin{subfigure}[t]{1\textwidth}
		\centering
		\includegraphics[scale=1]{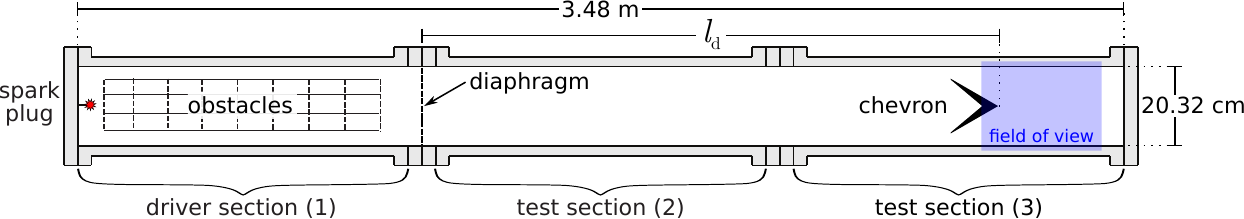}
		\caption{\label{fig:shock_tube:apparatus}Apparatus}
	\end{subfigure}
\\
	\begin{subfigure}[t]{0.66\textwidth}
		\centering
		\includegraphics[scale=1]{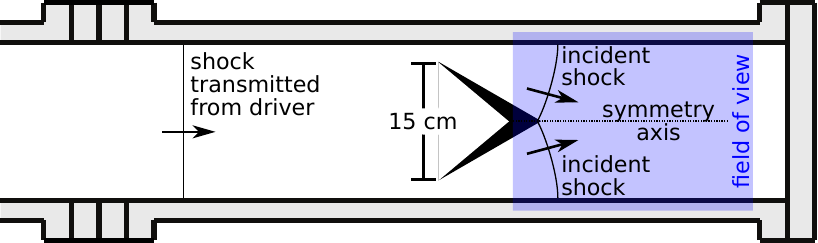}
		\caption{\label{fig:shock_tube:shock_transmission}Transmitted and incident shocks in the test section}
	\end{subfigure}%
	\begin{subfigure}[t]{0.33\textwidth}
		\centering
		\includegraphics[scale=1]{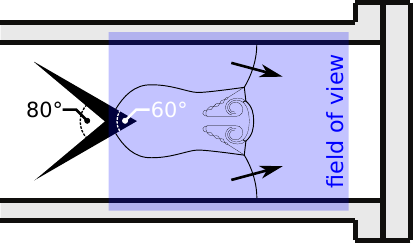}
		\caption{\label{fig:shock_tube:shoc_reflection}Subsequent shock reflection}
	\end{subfigure}
	\caption{\label{fig:shock_tube}Diagram of the shock tube with a depth of 1.91 cm; optical access throughout the third section; the chevron has an $80^{\circ}$ inner apex, $60^{\circ}$ outer apex, $15$ cm outside edges}
\end{figure}

Experiments were carried out in a detonation-driven aluminum shock tube illustrated in figure \ref{fig:shock_tube:apparatus} and described by \cite{maley_shock_2015}. The shock tube measured 3.48 m in length and had a rectangular cross-section measuring 20.32 cm tall by 1.91 cm in depth. A diaphragm separated the driver and test sections.

The shock tube was evacuated to $\lesssim 80$ Pa before gases were introduced. The driver section was filled with stoichiometric ethylene-oxygen ($\mathrm{C_2H_4 + 3 O_2}$). The test section was filled with a gas selected for its isentropic exponent, listed in table \ref{tab:experimental conditions}. Nitrogen was used for $\gamma_0=1.4$, an inert rich-propane-oxygen mixture ($\mathrm{0.8 C_3H_8 + 0.2 O_2}$) for $\gamma_0=1.15$, and normal hexane for $\gamma_0=1.06$. The subscript $0$ refers to the unshocked state.

Experiments were initiated with a spark 
that ignited the driver gas. A mesh of obstacles in the first third of the shock tube promoted detonation initiation in the driver section. The detonation ruptured the diaphragm, transmitting a shock into the test section.

The diaphragm was composed of aluminum foil covered by aluminum tape on the test gas side. An `\verb|I|' shaped slot matching the channel dimensions was cut through the tape. The foil separated the driver and test gases and opened along the slot in the tape when hit by the detonation. The tape kept the foil in place during opening, preventing diaphragm petals from polluting the test section with large fragments, though small fragments still formed. 
Experiment 10 used a plastic sheet as a diaphragm (an exception, from the diaphragm selection process).

The transmitted shock travelled through the test gas to a chevron-shaped obstacle. The top and bottom portions of the shock diffracted past the chevron (figure \ref{fig:shock_tube:shock_transmission}) and reflected from each other at the trailing edge (figure \ref{fig:shock_tube:shoc_reflection}). 
The chevron tip was located in the last third of the shock tube at a distance $\hat{l}_{\mathrm{d}}$ from the diaphragm, listed in table \ref{tab:experimental conditions}. The chevron limbs measured $15$ cm on the outside and had a $60^{\circ}$ outer apex. This made for reflection with a wedge angle of $\theta_{\mathrm{w}}=30^{\circ}$. The geometry was chosen to approximate the angle between shocks at the end of a detonation cell (\cite{strehlow_strength_1969,austin_role_2003,radulescu_origin_2009,bhattacharjee_experimental_2013}).

Glass walls on the last third of the shock tube allowed optical access. A Z-type schlieren with a vertical knife edge was used to capture high-speed videos with a Phantom v1210 camera at 77481 frames per second with 0.468 {\micro}s exposures and a resolution of 384$\times$288 pixels. The videos were centred downstream of the chevron with the trailing edge in view.
 
The strength of the shock interacting with the obstacle was controlled by the pressure ratio between driver and test gases. It was also affected by the distance between the diaphragm and chevron, and the diaphragm construction. Higher pressure ratios, shorter separation distances, and fewer layers of tape made for stronger shock waves.

\begin{table}
	\centering
	\caption{\label{tab:experimental conditions}Experimental conditions}
	\begin{tabular}{c|c|c|c|c||c|c|c|c|c|}
		\hline
		Experiment & Test mixture & $\hat{T}_0$ [$^{\circ}$C] & $\hat{p}_0$ [kPa] & $\hat{l}_{\mathrm{d}}$ [m] & $\hat{\lambda}_0$ [{\micro}m] & $\gamma_0$ & $M_{\mathrm{c}}$ & Figures & Movie
		\\ \hline
		1& $\mathrm{N_2}$ & 20.5 & 3.7 & 1.85 & 1.8 &  1.40 & 2.4 & \ref{fig:experiment:2017-04-20-A:13} to \ref{fig:experiment:2017-04-20-A:31} & 1
		\\
		2& $\mathrm{N_2}$ & 19.0 & 3.4 & 1.85 & 1.9 & 1.40 & 3.0 & \ref{fig:experiment:2017-05-04-D:13} to \ref{fig:experiment:2017-05-04-D:28} & 2
		\\
		3& $\mathrm{N_2}$ & 20.5 & 3.8 & 1.85 & 1.7 & 1.40 & 3.5 & \ref{fig:experiment:2017-04-20-C:10} to \ref{fig:experiment:2017-04-20-C:23} & 3
		\\
		4& $\mathrm{0.8 C_3H_8 + 0.2 O_2}$ & 21.0 & 6.96 & 1.85 & 0.407 & 1.15 & 2.4 & \ref{fig:experiment:2017-03-29-F:15} to \ref{fig:experiment:2017-03-29-F:39} & 4
		\\
		5& $\mathrm{0.8 C_3H_8 + 0.2 O_2}$ & 22.0 & 3.5 & 1.85 & 0.81 & 1.15 & 2.9 & \ref{fig:experiment:2017-04-06-B:14} to \ref{fig:experiment:2017-04-06-B:31} & 5
		\\
		6& $\mathrm{0.8 C_3H_8 + 0.2 O_2}$ & 21.5 & 3.9 & 1.85 & 0.74 & 1.15 & 3.5 & \ref{fig:experiment:2017-04-06-C:14} to \ref{fig:experiment:2017-04-06-C:27} & 6
		\\
		7& $\mathrm{C_6H_{14}}$ & 20.0 & 2.6 & 1.85 & 0.51 & 1.06 & 2.5 & \ref{fig:experiment:2017-03-29-B:16} to \ref{fig:experiment:2017-03-29-B:49} & 7
		\\
		8& $\mathrm{C_6H_{14}}$ & 22.5 & 2.6 & 1.85 & 0.52 & 1.06 & 2.7 & \ref{fig:experiment:2017-03-30-D:17} to \ref{fig:experiment:2017-03-30-D:42} & 8
		\\
		9& $\mathrm{C_6H_{14}}$ & 23.0 & 2.3 & 1.85 & 0.56 & 1.06 & 3.4 & \ref{fig:experiment:2017-03-30-G:15} to \ref{fig:experiment:2017-03-30-G:36} & 9
		\\
		10& $\mathrm{0.8 C_3H_8 + 0.2 O_2}$ & 23.0 & 14.3 & 0.32 & 0.199 & 1.15 & 4.0 & \ref{fig:experiment:2017-03-08-A:12} to \ref{fig:experiment:2017-03-08-A:29} & 10
		\\
		11& $\mathrm{C_6H_{14}}$ & 23.5 & $\sim$3.5 & 0.68 & 0.38 & 1.06 & 4.0 & \ref{fig:experiment:2017-03-22-D:15} to \ref{fig:experiment:2017-03-22-D:35} & 11
	\end{tabular}
\end{table}

\subsection{Experimental results}

Experimental results are shown in figures \ref{fig:experiments:g=1.4}, \ref{fig:experiments:g=1.15}, and \ref{fig:experiments:g=1.06}. They are a sample of the experiments performed. Videos are supplied as supplementary electronic material.

Each row of the figures shows three snapshots from one experiment. The first frame is taken immediately before reflection, and the two subsequent frames are taken from later times. The Mach number $M_{\mathrm{c}}$ noted in the figures is the shock strength on the chevron surface prior to reflection, calculated using the averaged shock spacing between frames, the inter-frame delay, and sound speed of the unshocked gas. The channel height of 0.2032 m is used to scale the photographs. Each row corresponds to a different experiment.

The schlieren photographs show horizontal density gradients. Bright features indicate an increase of density from left to right, and \textit{vice versa}. 

\subsubsection{Experiments in nitrogen ($\gamma_0=1.4$)}

\begin{figure}
	\centering
	\begin{subfigure}[t]{0.16\textwidth}
		\centering
		\adjincludegraphics[width=\textwidth,trim={0 0 {.6666\width} 0},clip]{{./figures/experiments/2017-04-20/A/13_annotated}.pdf}
		\caption{\label{fig:experiment:2017-04-20-A:13}$M_{\mathrm{c}}=2.4$}
	\end{subfigure}\hfill
	\begin{subfigure}[t]{0.32\textwidth}
		\centering
		\adjincludegraphics[width=\textwidth,trim={0 0 {.3333\width} 0},clip]{{./figures/experiments/2017-04-20/A/18_annotated}.pdf}
		\caption{\label{fig:experiment:2017-04-20-A:18}$Re = 1.9 \EXP{4}$, $\hat{t}=64.5$ {\micro}s}
	\end{subfigure}\hfill
	\begin{subfigure}[t]{0.48\textwidth}
		\centering
		\includegraphics[width=\textwidth]{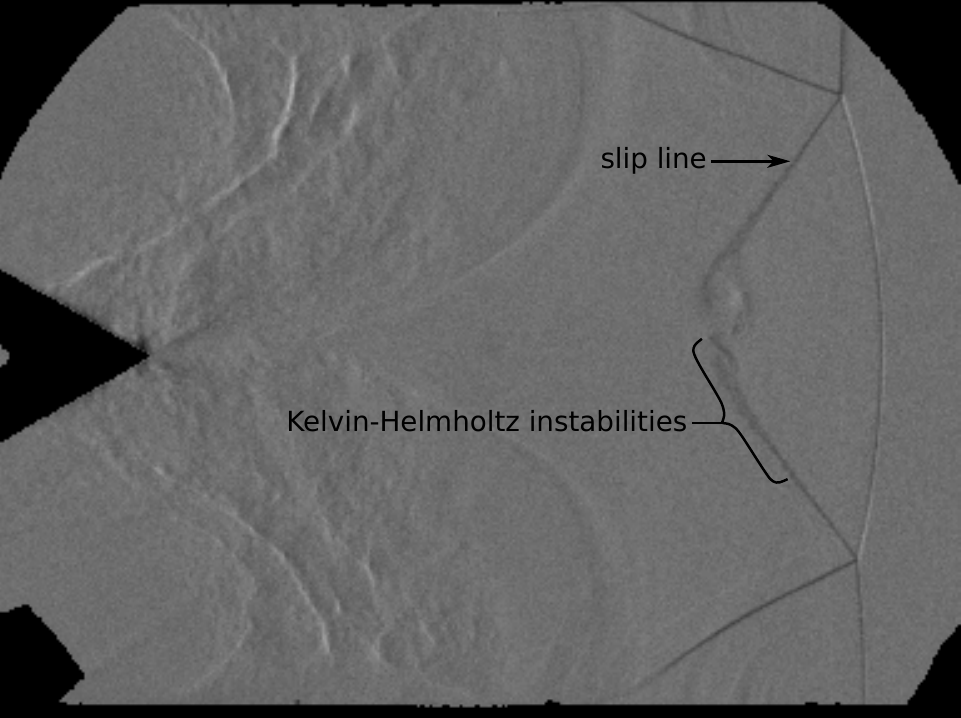}
		\caption{\label{fig:experiment:2017-04-20-A:31}$Re = 1.1 \EXP{5}$, $\hat{t}=232$ {\micro}s}
	\end{subfigure}
	\\
	\begin{subfigure}[t]{0.16\textwidth}
		\centering
		\adjincludegraphics[width=\textwidth,trim={0 0 {.6666\width} 0},clip]{{./figures/experiments/2017-05-04/D/13_annotated}.pdf}
		\caption{\label{fig:experiment:2017-05-04-D:13}$M_{\mathrm{c}}=3.0$}
	\end{subfigure}\hfill
	\begin{subfigure}[t]{0.32\textwidth}
		\centering
		\adjincludegraphics[width=\textwidth,trim={0 0 {.3333\width} 0},clip]{{./figures/experiments/2017-05-04/D/17_annotated}.pdf}
		\caption{\label{fig:experiment:2017-05-04-D:17}$Re = 2.3 \EXP{4}$, $\hat{t}=51.6\ \mu$s}
	\end{subfigure}\hfill
	\begin{subfigure}[t]{0.48\textwidth}
		\centering
		\includegraphics[width=\textwidth]{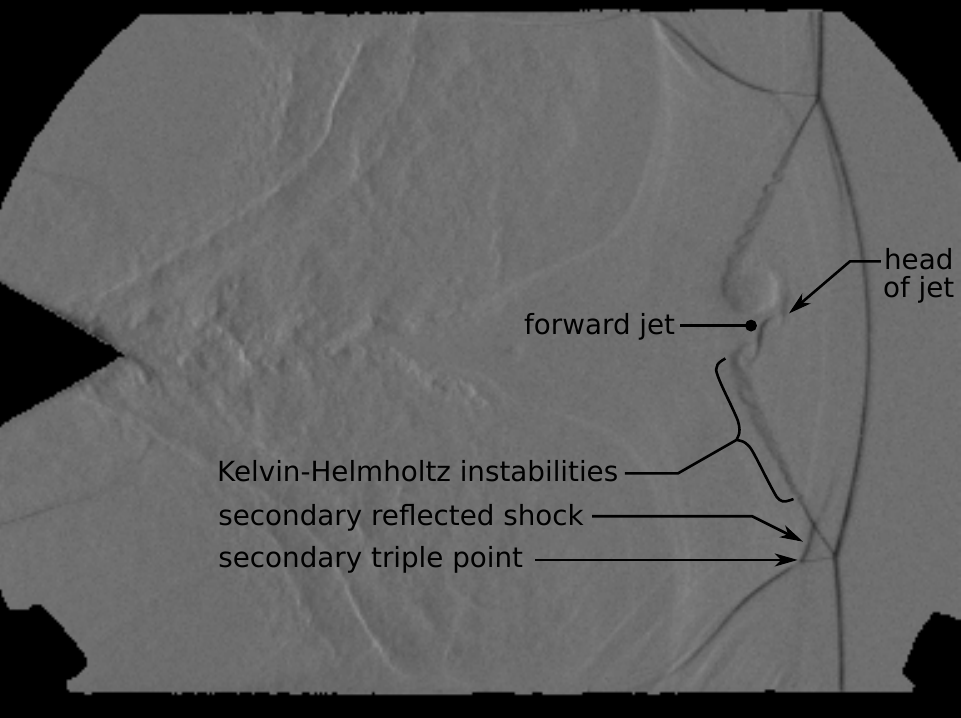}
		\caption{\label{fig:experiment:2017-05-04-D:28}$Re = 1.5 \EXP{5}$, $\hat{t}=194$ {\micro}s}
	\end{subfigure}
	\\
	\begin{subfigure}[t]{0.16\textwidth}
		\centering
		\adjincludegraphics[width=\textwidth,trim={0 0 {.6666\width} 0},clip]{{./figures/experiments/2017-04-20/C/10}.png}
		\caption{\label{fig:experiment:2017-04-20-C:10}$M_{\mathrm{c}}=3.5$}
	\end{subfigure}\hfill
	\begin{subfigure}[t]{0.32\textwidth}
		\centering
		\adjincludegraphics[width=\textwidth,trim={0 0 {.3333\width} 0},clip]{{./figures/experiments/2017-04-20/C/13}.png}
		\caption{\label{fig:experiment:2017-04-20-C:13}$Re = 2.5 \EXP{4}$, $\hat{t}=38.7$ {\micro}s}
	\end{subfigure}\hfill
	\begin{subfigure}[t]{0.48\textwidth}
		\centering
		\includegraphics[width=\textwidth]{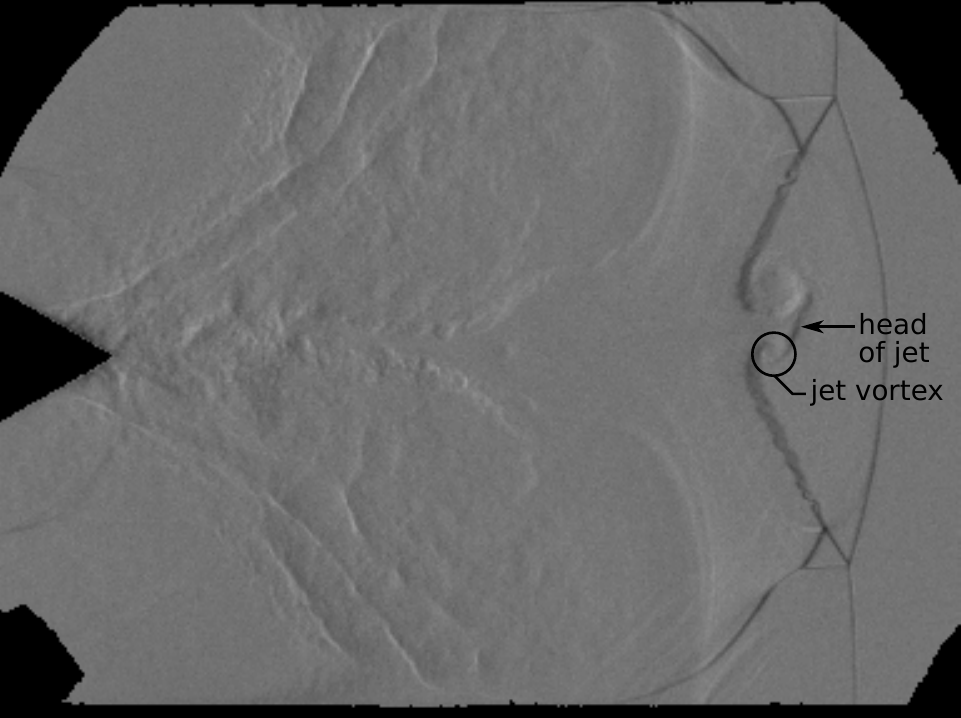}
		\caption{\label{fig:experiment:2017-04-20-C:23}$Re = 1.9 \EXP{5}$, $\hat{t}=168$ {\micro}s}
	\end{subfigure}
	\caption{\label{fig:experiments:g=1.4}Schlieren photographs of nitrogen experiments ($\gamma_0 = 1.4$, $\theta_{\mathrm{w}} = 30^{\circ}$)}
\end{figure}

Figure \ref{fig:experiment:2017-04-20-A:13} shows a schlieren photograph of an experiment in nitrogen gas, $\gamma_0=1.4$, immediately before shock reflection. The dark curves are two shocks that have diffracted over the obstacle, travelling to the right at $M_{\mathrm{c}}=2.4$. These form the incident shocks for the reflection in the following frames. The asymmetry in shock strengths prior to reflection was typically below $5 \% M_{\mathrm{c}}$. 
The flow field is shown $64.5$ {\micro}s later in figure \ref{fig:experiment:2017-04-20-A:18}. Single Mach reflections are formed with a nearly straight Mach stem. Slip lines emanate from the triple points and curl towards the Mach stem near the axis of symmetry. 
The flow field $232$ {\micro}s after reflection is shown in figure \ref{fig:experiment:2017-04-20-A:31}. The Mach stem has grown to occupy most of the channel and has curved due to unsteadiness. Kelvin-Helmholtz instability have appeared along the slip lines.

The Mach number is increased to $M_{\mathrm{c}}=3.0$ in the next experiment by increasing the pressure ratio between driver and test gas.  
A transitional Mach reflection is formed (figure \ref{fig:experiment:2017-05-04-D:17}) and transitions to a double Mach reflection (figure \ref{fig:experiment:2017-05-04-D:28}), seen by the formation of a secondary triple point and reflected shock where there was previously a kink. 
The slip line curls into a forward jet that is longer and closer to the Mach stem than before.

A double Mach reflection is formed (figure \ref{fig:experiment:2017-04-20-C:13}) in the next experiment, at $M_{\mathrm{c}}=3.5$, and vorticies begin to form at the head of the jet (figure \ref{fig:experiment:2017-04-20-C:23}).

\subsubsection{Experiments in propane with oxygen ($\gamma_0=1.15$)}

\begin{figure}
	\centering
	\begin{subfigure}[t]{0.16\textwidth}
		\centering
		\adjincludegraphics[width=\textwidth,trim={0 0 {.6666\width} 0},clip]{{./figures/experiments/2017-03-29/F/15}.png}
		\caption{\label{fig:experiment:2017-03-29-F:15}$M_{\mathrm{c}}=2.4$}
	\end{subfigure}\hfill
	\begin{subfigure}[t]{0.32\textwidth}
		\centering
		\adjincludegraphics[width=\textwidth,trim={0 0 {.3333\width} 0},clip]{{./figures/experiments/2017-03-29/F/22}.png}
		\caption{\label{fig:experiment:2017-03-29-F:22}$Re = 1.2 \EXP{5}$, $\hat{t}=90.3$ {\micro}s}
	\end{subfigure}\hfill
	\begin{subfigure}[t]{0.48\textwidth}
		\centering
		\includegraphics[width=\textwidth]{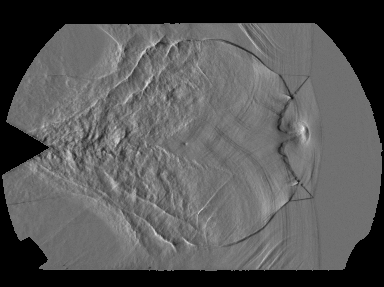}
		\caption{\label{fig:experiment:2017-03-29-F:39}$Re = 7.3  \EXP{5}$, $\hat{t}=310$ {\micro}s}
	\end{subfigure}
	\\
	\begin{subfigure}[t]{0.16\textwidth}
		\centering
		\adjincludegraphics[width=\textwidth,trim={0 0 {.6666\width} 0},clip]{{./figures/experiments/2017-04-06/B/14}.png}
		\caption{\label{fig:experiment:2017-04-06-B:14}$M_{\mathrm{c}}=2.9$}
	\end{subfigure}\hfill
	\begin{subfigure}[t]{0.32\textwidth}
		\centering
		\adjincludegraphics[width=\textwidth,trim={0 0 {.3333\width} 0},clip]{{./figures/experiments/2017-04-06/B/20}.png}
		\caption{\label{fig:experiment:2017-04-06-B:20}$Re = 7.9 \EXP{4}$, $\hat{t}=77.4$ {\micro}s}
	\end{subfigure}\hfill
	\begin{subfigure}[t]{0.48\textwidth}
		\centering
		\includegraphics[width=\textwidth]{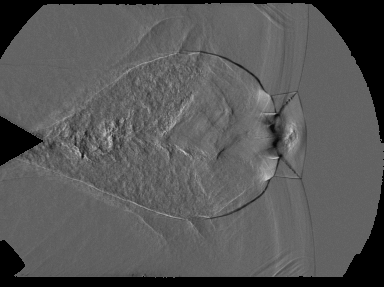}
		\caption{\label{fig:experiment:2017-04-06-B:31}$Re = 3.7 \EXP{5}$, $\hat{t}=219$ {\micro}s}
	\end{subfigure}
	\\
	\begin{subfigure}[t]{0.16\textwidth}
		\centering
		\adjincludegraphics[width=\textwidth,trim={0 0 {.6666\width} 0},clip]{{./figures/experiments/2017-04-06/C/14_annotated}.pdf}
		\caption{\label{fig:experiment:2017-04-06-C:14}$M_{\mathrm{c}}=3.5$}
	\end{subfigure}\hfill
	\begin{subfigure}[t]{0.32\textwidth}
		\centering
		\adjincludegraphics[width=\textwidth,trim={0 0 {.3333\width} 0},clip]{{./figures/experiments/2017-04-06/C/19}.png}
		\caption{\label{fig:experiment:2017-04-06-C:19}$Re = 7.9 \EXP{4}$, $\hat{t}=64.5$ {\micro}s}
	\end{subfigure}\hfill
	\begin{subfigure}[t]{0.48\textwidth}
		\centering
		\includegraphics[width=\textwidth]{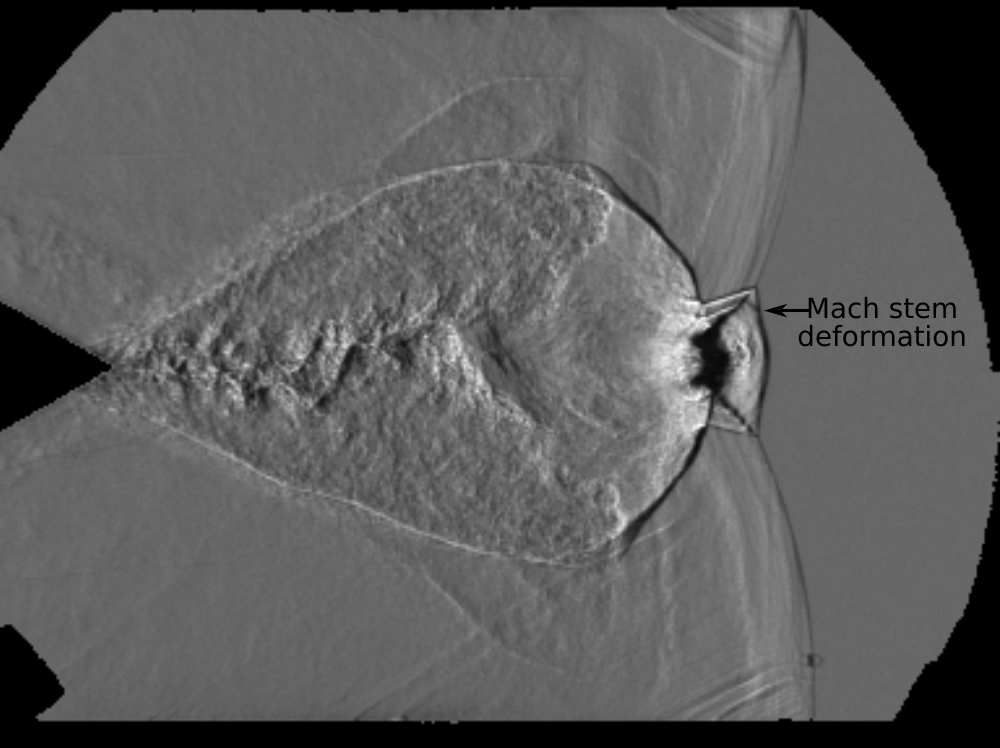}
		\caption{\label{fig:experiment:2017-04-06-C:27}$Re = 3.8 \EXP{5}$, $\hat{t}=168$ {\micro}s}
	\end{subfigure}
	\caption{\label{fig:experiments:g=1.15}Schlieren photographs of propane-oxygen experiments ($\gamma_0 = 1.15$, $\theta_{\mathrm{w}} = 30^{\circ}$)}
\end{figure}

The set of experiments in figure \ref{fig:experiments:g=1.15} are performed for the same Mach numbers but with a lower isentropic exponent of $\gamma_0=1.15$, by changing the test gas to rich-propane-oxygen.
Double Mach reflections are seen in all cases. 

At $M_{\mathrm{c}}=2.4$ (figure \ref{fig:experiment:2017-03-29-F:22}), the slip lines curl forward into a forward jet (figure \ref{fig:experiment:2017-03-29-F:39}) that is more distinct and closer to the Mach stem than all the $\gamma_0 = 1.4$ cases. The jet terminates with a vortex. 
Kelvin-Helmholtz instabilities on the slip line are entrained into the jet. 

At $M_{\mathrm{c}}=2.9$ (figure \ref{fig:experiment:2017-04-06-B:20}), the vortex at the head of the forward jet (figure \ref{fig:experiment:2017-04-06-B:31}) has a rough appearance caused by the entrainment of Kelvin-Helmholtz instabilities which create a turbulent substructure.

Increasing the Mach number to $M_{\mathrm{c}}=3.5$ causes triple points to form on the diffracting shocks (figure \ref{fig:experiment:2017-04-06-C:14}). 
These triple points are a result of the diffraction process (\cite{skews_shape_1967}). 
After reflection (figure \ref{fig:experiment:2017-04-06-C:27}), the forward jet impinges on the Mach stem, causing it to deform, bulge along the centre line, and possibly bifurcate. A large portion of the region behind the Mach stem has a turbulent appearance.

\subsubsection{Experiments in hexane ($\gamma_0=1.06$)}

\begin{figure}
	\centering
	\begin{subfigure}[t]{0.16\textwidth}
		\centering
		\adjincludegraphics[width=\textwidth,trim={0 0 {.6666\width} 0},clip]{{./figures/experiments/2017-03-29/B/16}.png}
		\caption{\label{fig:experiment:2017-03-29-B:16}$M_{\mathrm{c}}=2.5$}
	\end{subfigure}\hfill
	\begin{subfigure}[t]{0.32\textwidth}
		\centering
		\adjincludegraphics[width=\textwidth,trim={0 0 {.3333\width} 0},clip]{{./figures/experiments/2017-03-29/B/26}.png}
		\caption{\label{fig:experiment:2017-03-29-B:26}$Re = 1.1 \EXP{5}$, $\hat{t}=129$ {\micro}s}
	\end{subfigure}\hfill
	\begin{subfigure}[t]{0.48\textwidth}
		\centering
		\includegraphics[width=\textwidth]{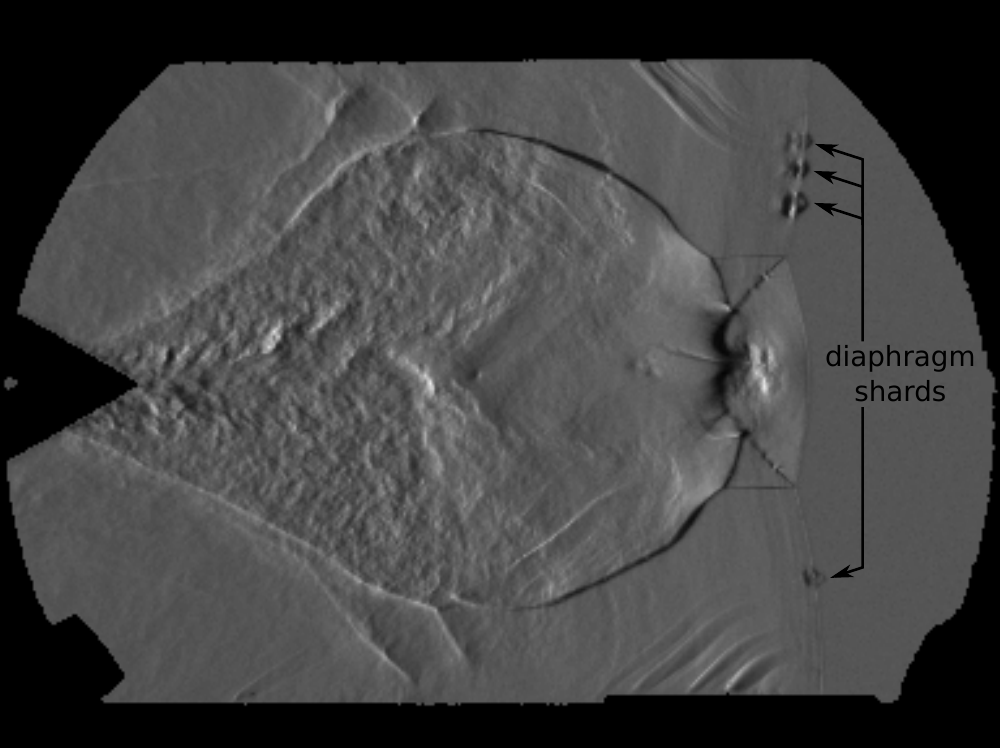}
		\caption{\label{fig:experiment:2017-03-29-B:49}$Re = 6.8 \EXP{5}$, $\hat{t}=426$ {\micro}s}
	\end{subfigure}
	\\
	\begin{subfigure}[t]{0.16\textwidth}
		\centering
		\adjincludegraphics[width=\textwidth,trim={0 0 {.6666\width} 0},clip]{{./figures/experiments/2017-03-30/D/17}.png}
		\caption{\label{fig:experiment:2017-03-30-D:17}$M_{\mathrm{c}}=2.7$}
	\end{subfigure}\hfill
	\begin{subfigure}[t]{0.32\textwidth}
		\centering
		\adjincludegraphics[width=\textwidth,trim={0 0 {.3333\width} 0},clip]{{./figures/experiments/2017-03-30/D/25}.png}
		\caption{\label{fig:experiment:2017-03-30-D:25}$Re = 5.1 \EXP{4}$, $\hat{t}=103$ {\micro}s}
	\end{subfigure}\hfill
	\begin{subfigure}[t]{0.48\textwidth}
		\centering
		\includegraphics[width=\textwidth]{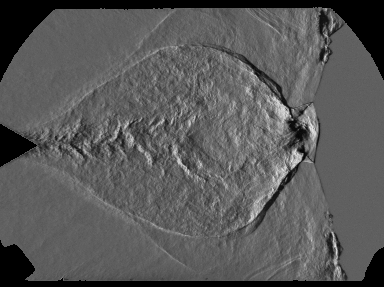}
		\caption{\label{fig:experiment:2017-03-30-D:42}$Re = 5.2 \EXP{5}$, $\hat{t}=323$ {\micro}s}
	\end{subfigure}
	\\
	\begin{subfigure}[t]{0.16\textwidth}
		\centering
		\adjincludegraphics[width=\textwidth,trim={0 0 {.6666\width} 0},clip]{{./figures/experiments/2017-03-30/G/15}.png}
		\caption{\label{fig:experiment:2017-03-30-G:15}$M_{\mathrm{c}}=3.4$}
	\end{subfigure}\hfill
	\begin{subfigure}[t]{0.32\textwidth}
		\centering
		\adjincludegraphics[width=\textwidth,trim={0 0 {.3333\width} 0},clip]{{./figures/experiments/2017-03-30/G/22_annotated}.pdf}
		\caption{\label{fig:experiment:2017-03-30-G:22}$Re = 1.0 \EXP{5}$, $\hat{t}=90.3$ {\micro}s}
	\end{subfigure}\hfill
	\begin{subfigure}[t]{0.48\textwidth}
		\centering
		\includegraphics[width=\textwidth]{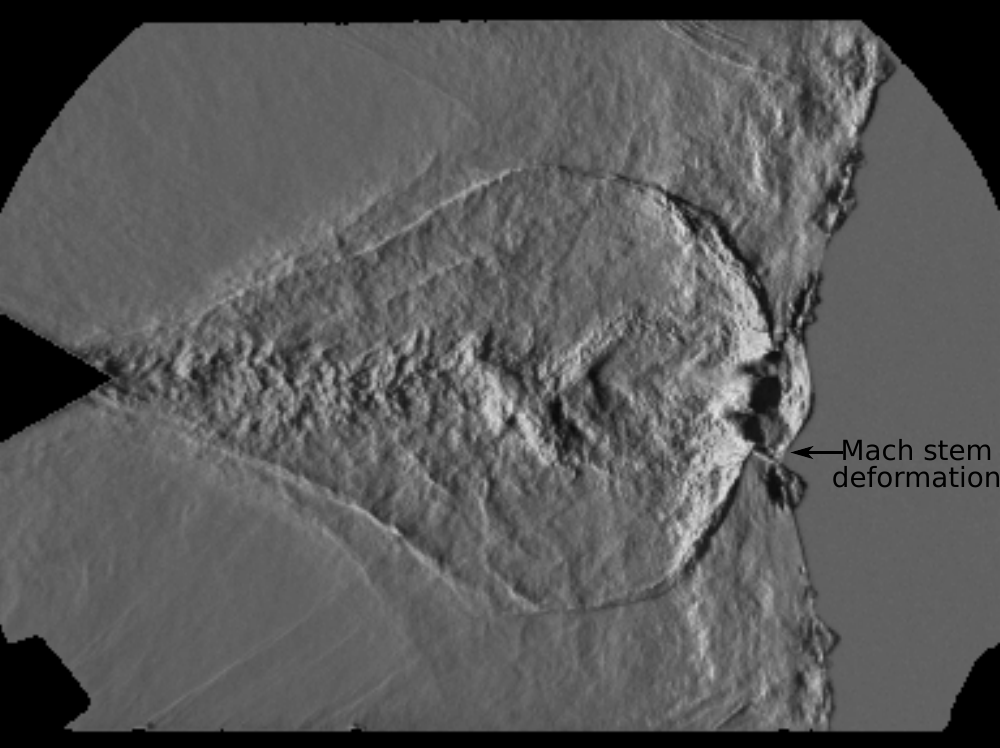}
		\caption{\label{fig:experiment:2017-03-30-G:36}$Re = 7.5 \EXP{5}$, $\hat{t}=258$ {\micro}s} 
	\end{subfigure}
	\caption{\label{fig:experiments:g=1.06}Schlieren photographs of hexane experiments ($\gamma_0 = 1.06$, $\theta_{\mathrm{w}} = 30^{\circ}$)}
\end{figure}

The isentropic exponent was lowered to $\gamma_0=1.06$ by employing n-hexane as the test gas. 
The double Mach reflection formed at $M_{\mathrm{c}}=2.5$ (figure \ref{fig:experiment:2017-03-29-B:49}) is qualitatively similar to the $M_{\mathrm{c}}=2.9$, $\gamma_0 = 1.15$ case (figure \ref{fig:experiment:2017-04-06-B:31}). Kelvin-Helmholtz instabilities appear close to the triple point and are entrained into the jet vortex, giving it a turbulent appearance that is distinct from the rest of the comparatively smooth-looking region behind the Mach stem. This rough, turbulent appearance of the vortex is present in all experiments where $M_{\mathrm{c}} \ge 2.5$ and $\gamma_0 = 1.06$.

The four triangular protrusions from the incident shocks seen in figure \ref{fig:experiment:2017-03-29-B:49} are small diaphragm shards that have overtaken the shock. This occurs because the highly compressible (low isentropic exponent) layer of gas between the shock and driver/test gas interface is thinner than in the other gases, which allows spall from the diaphragm to catch up to the shock front.

The Mach stem becomes shorter when the Mach number is increased to $M_{\mathrm{c}}=2.7$, making it difficult to see the flow field behind the shock front at early times (figure \ref{fig:experiment:2017-03-30-D:25}). A forward jet that impinges on the Mach stem becomes visible as the reflection grows (figure \ref{fig:experiment:2017-03-30-D:42}). The forward jet in the following experiment, $M_{\mathrm{c}}=3.4$, reaches the Mach stem and causes it to deform and possibly bifurcate (figure \ref{fig:experiment:2017-03-30-G:36}).

The schlieren photographs in the last two cases (figures \ref{fig:experiment:2017-03-30-D:17} to \ref{fig:experiment:2017-03-30-G:36}) have a textured appearance behind the Mach stem, incident and reflected shocks that is absent from the comparably smooth-looking flow fields seen all the previous cases, vortex aside. The textured appearance makes it difficult to observe fine features and suggests there is a flow instability present. This instability is not caused by the presence of driver gas, as will be discussed in sections \ref{sec:Experiments at higher Mach number}, \ref{sec:Shock instability at high Mach number and low isentropic exponent}, and the appendix.

\subsubsection{\label{sec:Experiments at higher Mach number}Experiments at a larger Mach number}

\begin{figure}
	\centering
	\begin{subfigure}[t]{0.16\textwidth}
		\centering
		\adjincludegraphics[width=\textwidth,trim={0 0 {.6666\width} 0},clip]{{./figures/experiments/2017-03-08/A/12}.png}
		\caption{\label{fig:experiment:2017-03-08-A:12}$\gamma_0 = 1.15$}
	\end{subfigure}\hfill
	\begin{subfigure}[t]{0.32\textwidth}
		\centering
		\adjincludegraphics[width=\textwidth,trim={0 0 {.3333\width} 0},clip]{{./figures/experiments/2017-03-08/A/20_annotated}.pdf}
		\caption{\label{fig:experiment:2017-03-08-A:20}$Re = 2.2 \EXP{6}$, $\hat{t}=103$ {\micro}s}
	\end{subfigure}\hfill
	\begin{subfigure}[t]{0.48\textwidth}
		\centering
		\includegraphics[width=\textwidth]{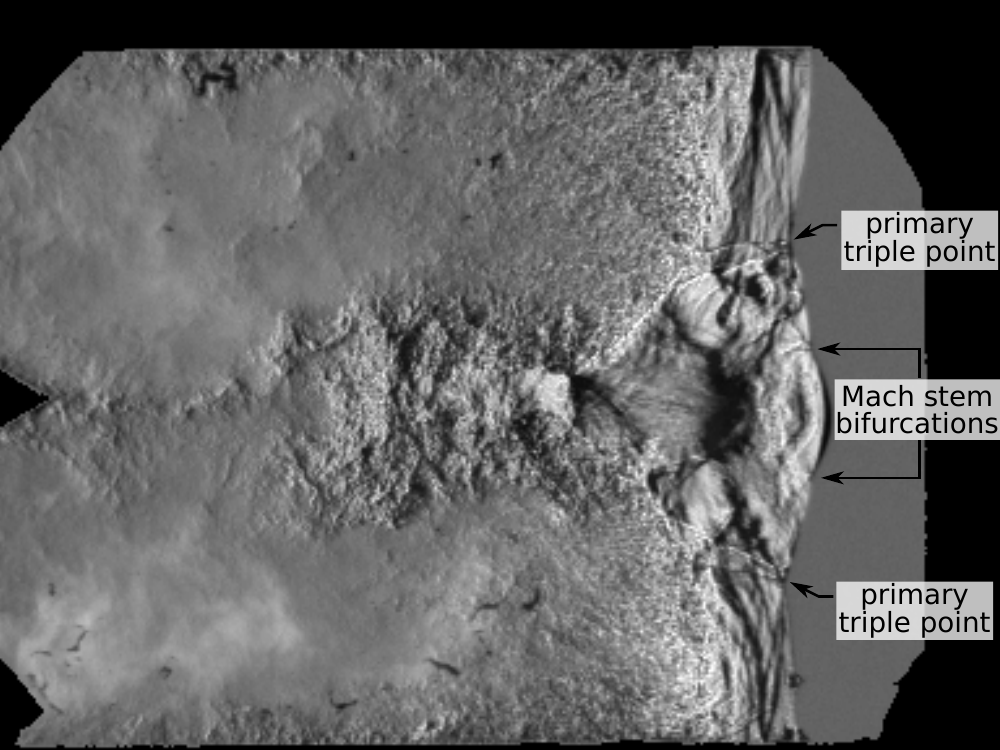}
		\caption{\label{fig:experiment:2017-03-08-A:29}$Re = 4.7 \EXP{6}$, $\hat{t}=219$ {\micro}s}
	\end{subfigure}
	\\
	\begin{subfigure}[t]{0.16\textwidth}
		\centering
		\adjincludegraphics[width=\textwidth,trim={0 0 {.6666\width} 0},clip]{{./figures/experiments/2017-03-22/D/15}.png}
		\caption{\label{fig:experiment:2017-03-22-D:15}$\gamma_0 = 1.06$}
	\end{subfigure}\hfill
	\begin{subfigure}[t]{0.32\textwidth}
		\centering
		\adjincludegraphics[width=\textwidth,trim={0 0 {.3333\width} 0},clip]{{./figures/experiments/2017-03-22/D/23_annotated}.png}
		\caption{\label{fig:experiment:2017-03-22-D:23}$Re = 5.8 \EXP{5}$, $\hat{t}=103$ {\micro}s}
	\end{subfigure}\hfill
	\begin{subfigure}[t]{0.48\textwidth}
		\centering
		\includegraphics[width=\textwidth]{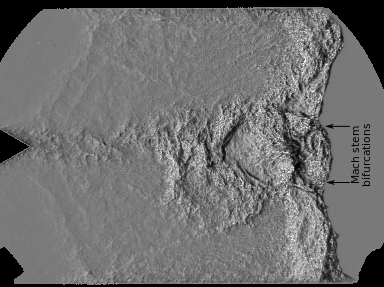}
		\caption{\label{fig:experiment:2017-03-22-D:35}$Re = 2.3 \EXP{6}$, $\hat{t}=258$ {\micro}s}
	\end{subfigure}
	\caption{\label{fig:experiments:Mc=4}Schlieren photographs of experiments at a larger Mach number ($M_{\mathrm{c}} = 4.0$, $\theta_{\mathrm{w}} = 30^{\circ}$; top row: propane-oxygen, bottom row: hexane)}
\end{figure}

In order to increase the Mach number further, the distance between the diaphragm and the chevron had to be shortened. 
The results are shown in figure \ref{fig:experiments:Mc=4}.

In both gases, double Mach reflections are formed with Mach stems that are taller than expected from the previous cases. This is due to the influence of the driver/test gas interface.
The forward jets reach the Mach stem and cause them to curve and bifurcate, seen in the last column of figure \ref{fig:experiments:Mc=4}.

Moving the diaphragm too close to the chevron causes the driver gas to interfere with the reflection. 
All of the previous experiments (figures \ref{fig:experiments:g=1.4} to \ref{fig:experiments:g=1.06}) are not affected by the presence of the driver gas. This is evidenced by the fact that their bow shocks (\textit{i.e.} the horizontal reflected shocks between the triple point and chevron, see figure \ref{fig:experiment:2017-03-30-G:22}) remain intact and attached to the chevron. These shocks are absent in figure \ref{fig:experiments:Mc=4} where contamination by the high sound speed driver gas causes them to disperse. 
Simulations in the appendix support that the driver/test gas interface is far from the leading shock in the previous experiments.

\subsubsection{Summary of experimental results}

A series of experiments were performed with $Re \sim 10^{5}$ to $10^{6}$, $M_{\mathrm{c}} = 2.4$ to $4$, and $\gamma_0 = 1.4$ to $1.06$.
The experiments show that decreasing the isentropic exponent and increasing the Mach number causes a transition from single Mach reflection, to transitional Mach reflection, onto double Mach reflection. The forward jet and vortex gradually become stronger and eventually deform the Mach stem as the isentropic exponent decreases and the Mach number is increased. Concurrently, Kelvin-Helmholtz instability become entrained in the vortex, resulting in a turbulent structure behind the Mach stem in the cases with lower isentropic exponents.

\subsubsection{\label{sec:Experiment Reynolds numbers}Mach stem bulging}

\begin{figure}
	\centering
	\begin{subfigure}[t]{0.48\textwidth}
		\centering
		\includegraphics[width=\textwidth]{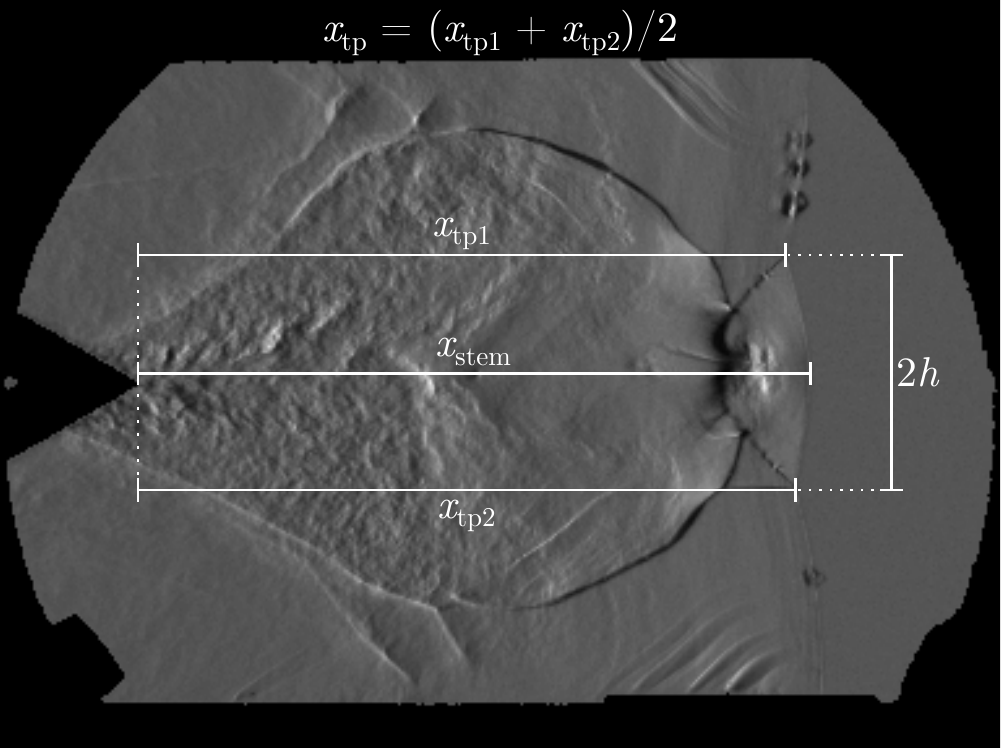}
	\end{subfigure}
	\caption{\label{fig:experiments:bulging_calculation}Definitions of $x_{\mathrm{tp}}$, $x_{\mathrm{stem}}$ and $h$}
\end{figure}

\begin{figure}
	\centering
	\begin{subfigure}[b]{0.33\textwidth}
		\centering
		\includegraphics[scale=1]{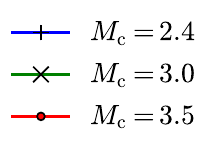}
		\\
		\includegraphics[scale=1]{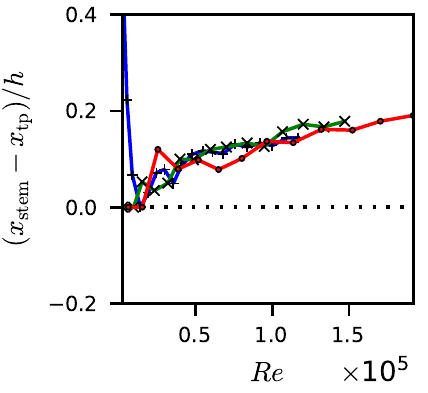}
		\caption{$\gamma_0=1.4$}
	\end{subfigure}%
	\begin{subfigure}[b]{0.33\textwidth}
		\centering
		\includegraphics[scale=1]{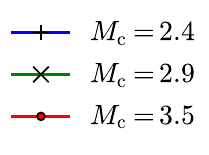}
		\\
		\includegraphics[scale=1]{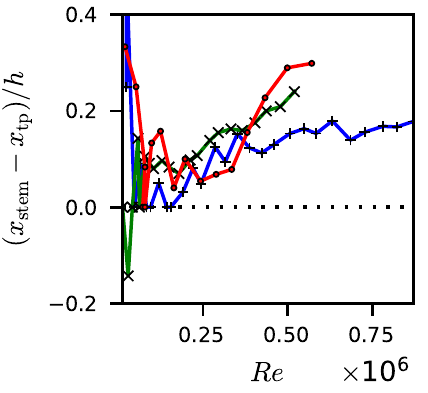}
		\caption{$\gamma_0=1.15$}
	\end{subfigure}%
	\begin{subfigure}[b]{0.33\textwidth}
		\centering
		\includegraphics[scale=1]{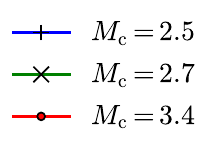}
		\\
		\includegraphics[scale=1]{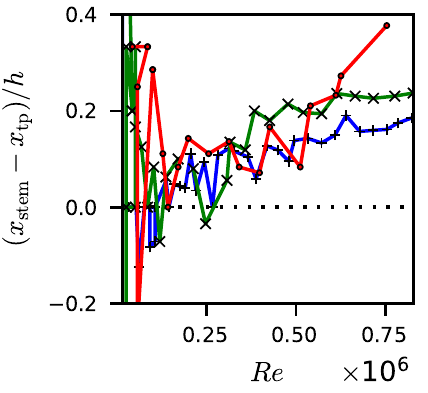}
		\caption{\label{fig:experiment:quantitative:g=1.06}$\gamma_0=1.06$}
	\end{subfigure}%
	\caption{\label{fig:experiment:quantitative}Evolution of the Mach stem bulging in experiments}
\end{figure}

The size or `strength' of the forward jet and vortex is difficult to measure at moderate Mach numbers ($ \gtrsim 3$) and low isentropic exponents ($\le 1.15$) due to the short Mach stems and the rough appearance of the photos behind the shock front. The bulging of the Mach stem foot is measured instead, as a proxy of the jet's strength.
Bulging is measured as the horizontal distance between the right-most point of the Mach stem ($x_{\mathrm{stem}}$) and the triple point ($x_{\mathrm{tp}}$, averaged between the top and bottom triple points), and normalized by the Mach stem height $h$ (see figure \ref{fig:experiments:bulging_calculation}). Bulging, $(x_{\mathrm{stem}} - x_{\mathrm{tp}}) / h$, is plotted against the Reynolds number in figure \ref{fig:experiment:quantitative}. Each point corresponds to one frame from the videos.

The Reynolds number for a Mach reflection was defined as
\begin{equation}
\label{eq:experiment:Reynolds number}
Re = \frac{u_{\mathrm{shear}}}{\nu_{\mathrm{mean}}} h
\end{equation}
by \cite{rikanati_shock-wave_2006,rikanati_secondary_2009} to study  Kelvin-Helmholtz instability along the slip line. Here $u_{\mathrm{shear}}$ is the velocity difference across the slip line and $\nu_{\mathrm{mean}}$ is the kinematic viscosity averaged across the slip line.
This definition takes into account the velocity and viscosity across the slip line (which forms the forward jet) where diffusion is expected to be important, and the Mach stem height $h$ provides an easily measured characteristic length scale.

Three shock theory is used to get analytical estimates of the shear velocity and viscosity, which are assumed constant. The inputs required for the calculation are the wedge angle $\theta_{\mathrm{w}}=30^{\circ}$, isentropic exponent $\gamma_0$, incident shock strength $M_{\mathrm{c}}$, and the unshocked temperature and pressure listed in table \ref{tab:experimental conditions}.
The estimates for ${\hat{u}_{\mathrm{shear}} / \hat{\nu}_{\mathrm{mean}} = Re / \hat{h}}$ are listed in table \ref{tab:experimental:Reynolds number}, giving  Reynolds numbers as a function of Mach stem height. 
The Mach stem height is measured directly from experiments as half the vertical distance between triple points. For example, the distance between triple points in figure \ref{fig:experiment:2017-04-20-A:18} is $2\hat{h} = 24 $ mm, and $ Re / \hat{h} = 1602$ mm$^{-1}$ (table \ref{tab:experimental:Reynolds number}), yielding a Reynolds number of $Re = 19000$.

\begin{table}
	\centering
	\caption{\label{tab:experimental:Reynolds number}Estimate of Reynolds number growth rates}
	\begin{tabular}{cc|c|c|c|c|c|c|c|c|c|c|c}
		\hline
		$\gamma_0$ & &\multicolumn{3}{|c|}{1.4} & \multicolumn{4}{|c|}{1.15} & \multicolumn{4}{|c}{1.06} 
		\\ 
		$M_{\mathrm{c}}$ & & 2.4 & 3.0 & 3.5 & 2.4 & 2.9 & 3.5 & 4.0 & 2.5 & 2.7 & 3.4 & 4.0
		\\ 
		$Re / \hat{h}$ & $[\mathrm{mm}^{-1}]$ & 1602 & 2138 & 2797 & 14490 & 11171 & 17635 & 79818 & 18523 & 22926 & 38219 & 78506
	\end{tabular}
\end{table}

In a self-similar pseudo-steady reflection, Mach stem bulging would draw a horizontal line in figure \ref{fig:experiment:quantitative}. However, the incident shocks are curved in these experiments, causing the reflection angle and shock strength to change continuously, leading to more bulging as Reynolds number increases.

Bulging is approximately equal in all gases at $M_{\mathrm{c}} \approx 2.5$ and is independent of Mach number in the $\gamma_0=1.4$ experiments because the jet never reaches the shock front. The additional bulging in the $\gamma_0=1.15$ and $\gamma_0=1.06$ gases at higher Mach numbers is caused when the forward jet approaches or contacts the Mach stem. 
Jetting becomes strong enough to cause bulging when the isentropic exponent is sufficiently low ($\gamma_0 \le 1.15$) and the Mach number is sufficiently large ($M_{\mathrm{c}} \gtrsim 3$). The vortex is turbulent in these cases.

\section{Numerical prediction}\label{sec:numerical prediction}

The experiments were limited to moderate Mach numbers and large Reynolds numbers. Numerical simulations will now be used to elucidate the flow field at early stages of the reflection and at higher Mach numbers. The numerical study is an extension of the work by \cite{lau-chapdelaine_viscous_2016}.

\subsection{Numerical technique}\label{sec:numerical technique}

The two-dimensional, laminar, unsteady Navier-Stokes equations for a calorically perfect gas were used. 
A Prandtl number of $3/4$ was used for all the simulations and the isentropic exponent was uniform throughout the domain.
Bulk viscosity was neglected, and viscosity and heat conductivity were assumed to be power laws of temperature and a known reference state 
\begin{equation}
\label{eq:diffusion_coefficients}
\hat{\mu} = \hat{\mu}_{\mathrm{ref}} \sqrt{\frac{\hat{T}}{\hat{T}_{\mathrm{ref}}}} \mathrm{\ \ \ \ and\ \ \ \ } \hat{k} = \hat{k}_{\mathrm{ref}} \sqrt{\frac{\hat{T}}{\hat{T}_{\mathrm{ref}}}} .
\end{equation}
The equations were non-dimensionalized using the unshocked gas (subscript 0) as reference state (subscript ``ref'')
\begin{equation}
\label{eq:non-dimensionalization}
\hat{\rho}_{\mathrm{ref}} = \hat{\rho}_{0} \mathrm{,\ \ \ \ } 
\hat{p}_{\mathrm{ref}} = \hat{p}_{0} \mathrm{,\ \ \ \ }
\hat{T}_{\mathrm{ref}} = \hat{T}_{0} \mathrm{,\ \ \ \ }
\hat{x}_{\mathrm{ref}}= \hat{\lambda}_{0} = \frac{\hat{\mu}_{0}}{\hat{p}_{0}} \sqrt{\frac{\pi \hat{R}_{\mathrm{s}} \hat{T}_{0}}{2}}
\end{equation}
where $\rho$ is density, $p$ is pressure, $T$ is temperature, $x$ is the spatial dimension, $R_{\mathrm{s}}$ is the specific gas constant, and the circumflex (hat) indicates dimensional variables. The spatial dimension was non-dimensionalized by the mean free path in the unshocked gas $\hat{\lambda}_0$, listed in table \ref{tab:experimental conditions} for the experimental conditions, and time was non-dimensionalized by $\hat{t}_{\mathrm{ref}} = \hat{\lambda}_0 \sqrt{\hat{\rho}_0/\hat{p}_0}$ for consistency.

{\color{black}The Navier-Stokes equations are valid in the continuum regime (Knudsen number $Kn = \lambda / L < 0.01$, $L$ is a characteristic length scale), and in the slip flow regime ($0.01 < Kn < 0.1$), where special conditions are required near walls (\cite{faghriTransportPhenomenaMultiphase2006}). Because shock reflection from a symmetry condition is studied instead of reflection from a wall, the Navier-Stokes equations remain valid in the slip flow regime. The simulations that will be presented in the next section (figure \ref{fig:simulations:gamma=1.06,Mi,Re},  at $Re \approx 1000$) have Knudsen numbers of  $\lambda_0 / h = 0.009$ to $0.06$ in the unshocked gas and $\lambda_{\mathrm{M}} / h = 0.001$ to $0.003$ behind the Mach shock, where the slip line and forward jet lie, which are well within the validity range of the Navier-Stokes equations.}

The equations were solved using the \verb|mg| software package developed by Falle (\cite{falle_self-similar_1991,falle_upwind_1996}). A second-order Godunov scheme using a monotonized central symmetric flux limiter (\cite{van_leer_towards_1977}) solved convective terms, and diffusion was solved explicitly in time with second order central differences in space.

\begin{figure}
	\centering
	\begin{subfigure}[t]{0.5\textwidth}
		\centering
		\includegraphics[scale=0.5]{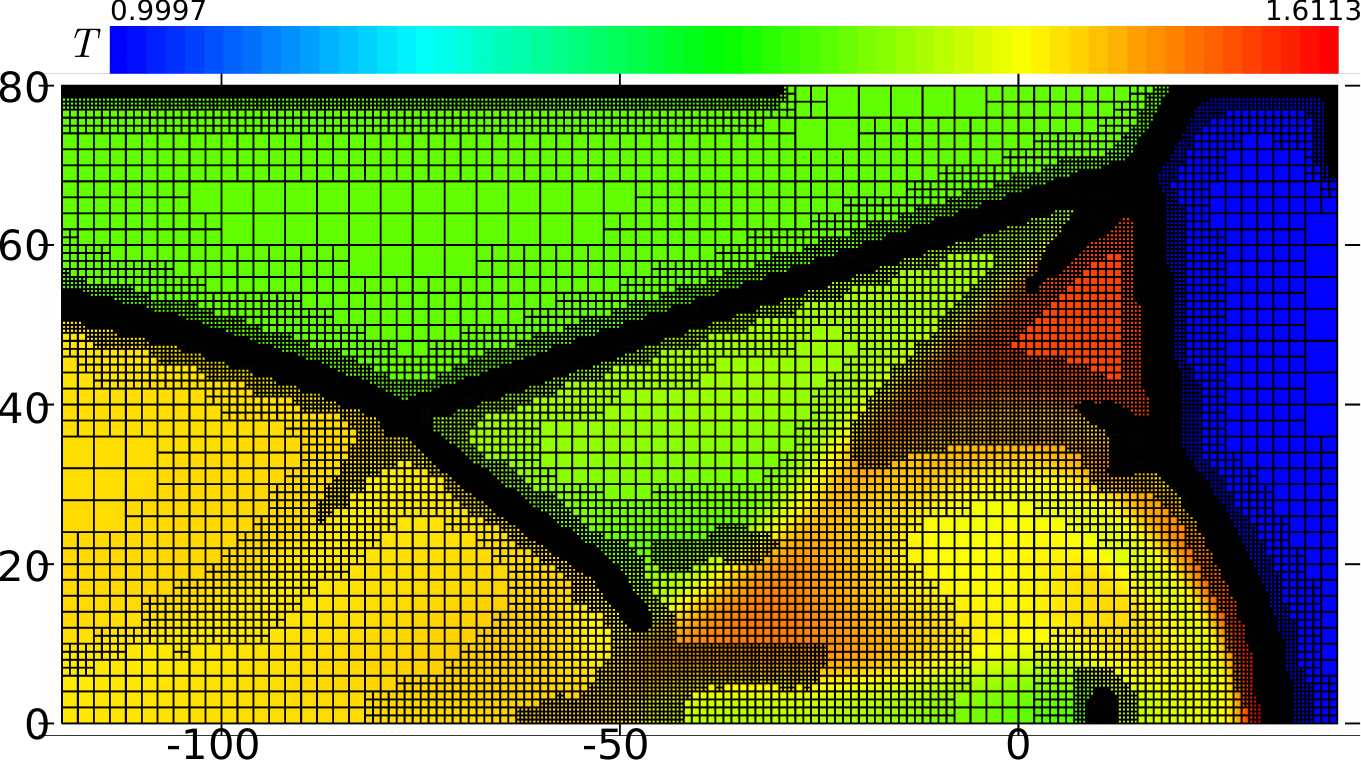}
		\caption{Viscous}
	\end{subfigure}%
	\begin{subfigure}[t]{0.5\textwidth}
		\centering
		\includegraphics[scale=0.5]{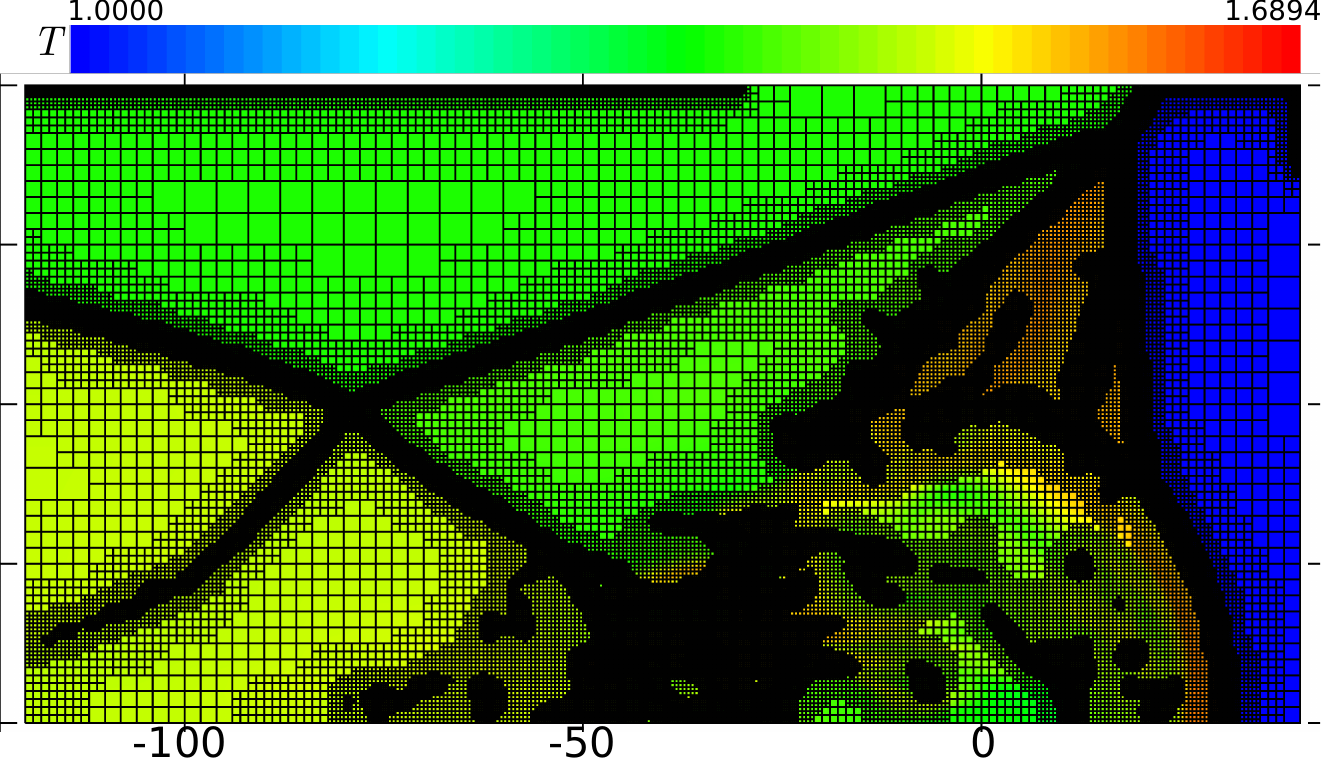}
		\caption{Inviscid}
	\end{subfigure}%
	\caption{\label{fig:AMR}Adaptive computational mesh overlayed on a temperature plot ($M=3.5$, $\gamma=1.06$, $t=132$)}
\end{figure}

\begin{figure}
	\centering
	\begin{subfigure}[t]{1\textwidth}
		\centering
		\includegraphics[scale=1]{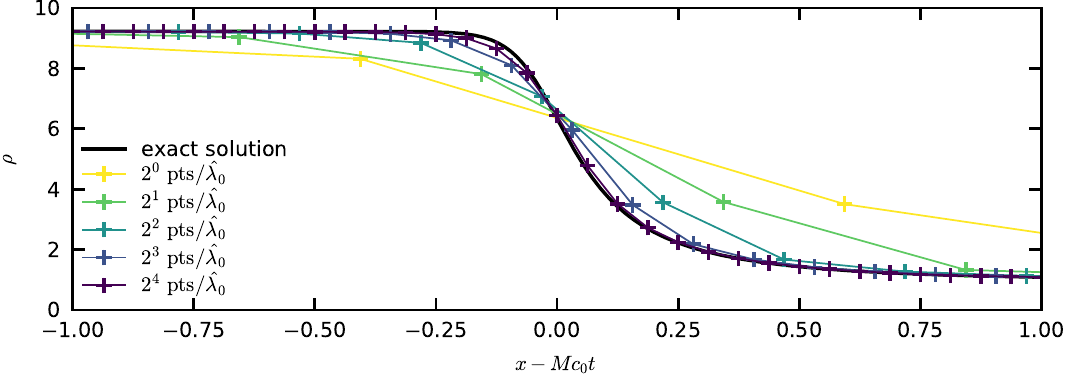}
	\end{subfigure}%
	\caption{\label{fig:shock_profile}Comparison of shock profiles evolved from a discontinuity to $t=4$ at different resolutions and the steady solution of the Navier-Stokes equations ($M=3.5$, $\gamma=1.06$)}
\end{figure}

A Cartesian grid with adaptive mesh refinement (\cite{sharpeNumericalSimulationsPremixed2011}) was used. The mesh was refined by a factor of two in both directions whenever a relative tolerance of 1\% was exceeded between existing mesh levels for density, pressure, or velocity. The refinement was extended five to ten cells in all directions from the cell needing refinement. Figure \ref{fig:AMR} shows the resulting adaptive mesh in viscous and inviscid simulations.
A maximal resolution of 16 points per mean free path in the unshocked gas was used. Under the conditions of experiment 9, for example, the mean free path of the test gas was 0.56 {\micro}m (before being shocked). At maximum resolution, the distance of one mean free path is covered by up to 16 grid points, resulting in a maximum resolution of 16 grid points per 0.56 {\micro}m, or 0.035 {\micro}m between every grid point. Figure \ref{fig:shock_profile} shows this resolution is sufficient to accurately reproduce shock wave thickness.

{\color{black}The Navier-Stokes shock structure recovers the experimental and direct simulation Monte-Carlo (DSMC) results at low Mach numbers but deteriorates at moderate and higher Mach numbers (\cite{uribeShockwaveStructureBased2018}). The lack of phenomenological models that can recreate the strong shock structure remains an open problem. The purpose of using shock-resolved Navier-Stokes simulations is to remove the solution's dependence on solver type and to ensure the correct Navier-Stokes solution is reached for all larger features, such as the slip line, forward jet, vortex, and their interaction with the shock front. It also provides properly resolved results for future shock reflection studies which may seek to compare other models or simulations with reduced resolutions. Simulating shock reflection using DSMC methods would be a further improvement, but these are even more computationally demanding.}
The Navier-Stokes simulations require much computational time due to the high resolution and the viscous Courant-Friedrichs-Lewy (CFL) criterion. As a result, the viscous simulations were limited to $\gamma=1.06$. Reflections at this isentropic exponent have short Mach stems which permits the use of small domains, and their Reynolds numbers grow quickly thus necessitating fewer time steps. The maximum time step size was determined using the CFL condition with $\mathrm{CFL} = 0.4$ on the refined meshes.

\begin{figure}
	\centering
	\begin{subfigure}[t]{0.5\textwidth}
		\centering
		\includegraphics[scale=1]{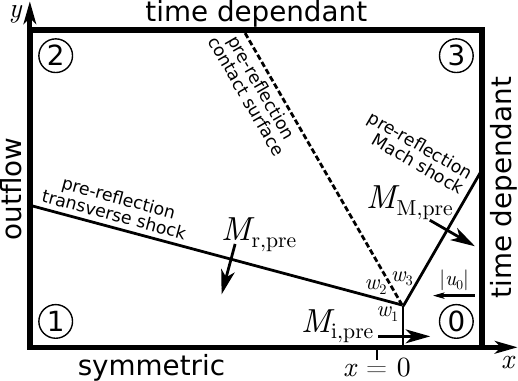}
		\caption{\label{fig:initial_conditions:a}$t=-1$: initial and boundary conditions}
	\end{subfigure}\hfill%
	\begin{subfigure}[t]{0.5\textwidth}
		\centering
		\includegraphics[scale=1]{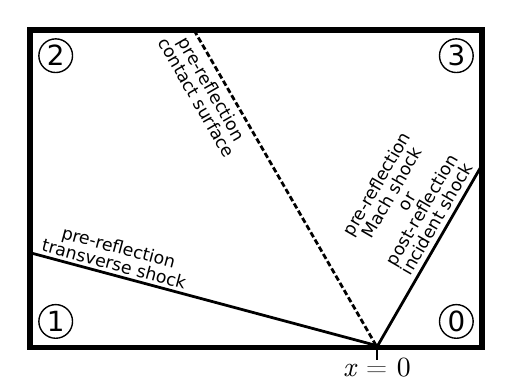}
		\caption{\label{fig:initial_conditions:b}$t = 0$: triple point reaches reflecting surface}
	\end{subfigure}%
	\\
	\begin{subfigure}[t]{0.5\textwidth}
		\centering
		\includegraphics[scale=1]{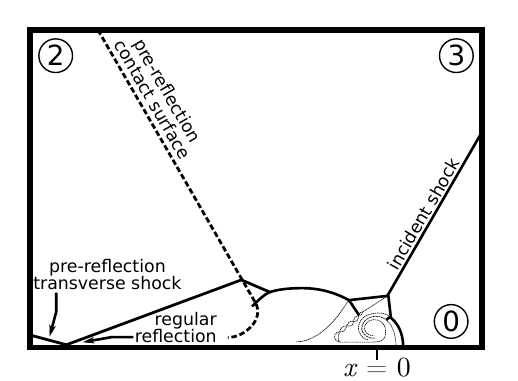}
		\caption{\label{fig:initial_conditions:c}$t > 0$: Mach reflection forms and grows}
	\end{subfigure}\hfill%
	\begin{subfigure}[t]{0.5\textwidth}
		\centering
		\includegraphics[scale=1]{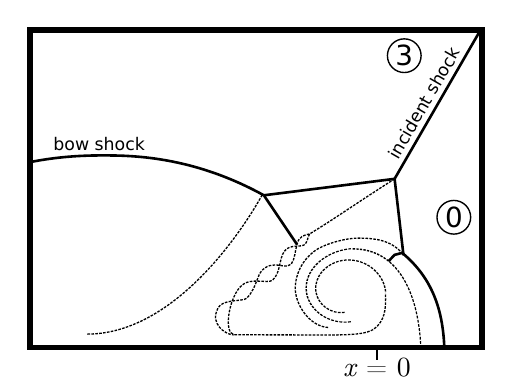}
		\caption{\label{fig:initial_conditions:d}$t \gg 0$: pre-reflection features leave domain}
	\end{subfigure}%
	\caption{\label{fig:initial_conditions}Initial and boundary conditions of simulations, and triple shock reflection; \raisebox{.5pt}{\textcircled{\raisebox{-1pt} {0}}} unshocked state, \raisebox{.5pt}{\textcircled{\raisebox{-1pt} {1}}} pre-reflection incident shock state, \raisebox{.5pt}{\textcircled{\raisebox{-1pt} {2}}} pre-reflection transverse shock state, \raisebox{.5pt}{\textcircled{\raisebox{-1pt} {3}}} pre-reflection Mach shock state and post-reflection incident shock state}
\end{figure}

The initial conditions consisted of a triple-shock, calculated using the ideal three shock theory, whose triple point was imposed above a symmetry boundary, as illustrated in figure \ref{fig:initial_conditions:a}. The large arrows point from the post-shock state to the pre-shock state. The initial conditions can be recreated using the data from table \ref{tab:simulation conditions} by using the shock Mach numbers with the Rankine-Hugoniot equations, the angles $w$ between discontinuities, and the reference frame speed $u_0$  (post-reflection Mach stem speed, calculated using three shock theory). Using a triple shock as initial condition does away with non-physical leaky boundary conditions and provides an unambiguous way to pose the problem (\cite{lau-chapdelaine_viscous_2016}). It also closely resembles triple shock reflections in detonations. 

\begin{table}
	\centering
	\caption{\label{tab:simulation conditions}Initial conditions for triple shock simulations, see figure \ref{fig:initial_conditions}; pre-reflection shock Mach numbers are listed so $M_{\mathrm{i}} = M_{\mathrm{M,pre}}$;  $\gamma=1.06$, $v_0=0$, $p_0=1$, $\rho_0=1$, $w_0 = 150^{\circ}$}
	\begin{tabular}{c|c|c|c|c|c|c}
		\hline
		$M_{\mathrm{M,pre}}$ & $M_{\mathrm{i,pre}}$ & $M_{\mathrm{r,pre}}$ & $u_0$ & $w_1$ & $w_2$ & $w_3$ 
		\\ \hline
		2.5 & 2.03663 & 1.22304 & -3.14027 & 104.462$^{\circ}$ & 44.6553$^{\circ}$ & 60.8829$^{\circ}$ 
		\\
		3 & 2.45729 & 1.21608 & -3.74700 & 118.124$^{\circ}$ & 36.2915$^{\circ}$ & 55.585$^{\circ}$ 
		\\
		3.5 & 2.45762 & 1.21592 & -3.74647 & 118.108$^{\circ}$ & 36.2434$^{\circ}$ & 55.6491$^{\circ}$ 
		\\
		4 & 3.30702 & 1.20460 & -4.95286 & 140.018$^{\circ}$ & 24.1620$^{\circ}$ & 45.8199$^{\circ}$ 
		\\
		5 & 4.16138 & 1.19658 & -6.15981 & 141.205$^{\circ}$ & 23.5999$^{\circ}$ & 45.1955$^{\circ}$ 
		\\
		6 & 5.01458 & 1.19159 & -7.36482 & 145.904$^{\circ}$ & 21.2707$^{\circ}$ & 42.8248$^{\circ}$ 
	\end{tabular}
\end{table}

In this configuration, the triple point initially moves downwards and to the left (figure \ref{fig:initial_conditions:a} to \ref{fig:initial_conditions:b}) because the pre-reflection incident shock is slower than $u_0$. The initial distance of the triple point above the symmetry boundary was chosen to allow the viscous shock structures to develop before reflection. The triple point reaches the origin on the symmetry boundary at $t=0$ (figure \ref{fig:initial_conditions:b}) and reflects. The pre-reflection Mach stem (the oblique shock between states \raisebox{.5pt}{\textcircled{\raisebox{-1pt} {0}}} and \raisebox{.5pt}{\textcircled{\raisebox{-1pt} {3}}}, figure \ref{fig:initial_conditions:b}) becomes the post-reflection incident shock and a Mach reflection is formed (figure \ref{fig:initial_conditions:c}). The pre-reflection slip line, incident and transverse shocks are washed out to the left because they travel slower than the frame of reference (figure \ref{fig:initial_conditions:c} to \ref{fig:initial_conditions:d}).

The top and right boundaries were functions of time, moving the initial shock and slip line states along the boundaries. An outflow condition with zero normal gradient was used on the left boundary. The domain was sized to fit the double Mach reflection structure at the target simulation time. The remaining parts of the reflection (\textit{e.g.} bow shock, figure \ref{fig:initial_conditions:d}) were allowed to flow out of the domain. Comparison of different domain sizes showed the domains chosen were sufficiently large to have no effect on the phenomena being studied.

The target simulation time was found for a desired Reynolds number of $Re_{\mathrm{target}}=1000$, defined in equation \ref{eq:experiment:Reynolds number}.
The Mach stem height was estimated \textit{a priori} from three shock theory as $h = \tan(\chi) M_{\mathrm{M}} c_0 t$, with post-reflection triple point path angle  $\chi$, post-reflection Mach stem Mach number $M_{\mathrm{M}}$, and unshocked sound speed $c_0$. This yields an expression for the target time
\begin{equation}
\label{eq:target_time}
	t_{\mathrm{target}} = \frac{\nu_{\mathrm{mean}}} {u_{\mathrm{shear}} \tan(\chi) M_{\mathrm{M}} c_0} Re_{\mathrm{target}},
\end{equation}
where $\nu_{\mathrm{mean}}$ and $u_{\mathrm{shear}}$ are also estimated from three shock theory.
The solution was exported at fixed time steps throughout the simulation. Reaching the target time for $M_{\mathrm{i}}=2.5$ in a domain measuring 352$\times$192 $\lambda_{0}$ took approximately 20 days on 60 AMD Opteron 6282 processor cores with a coarse grid of 22$\times$12 cells and 8 levels of refinement, for a finest possible grid of 5632$\times$3072 cells. The computational resources limited the scope of simulations to $Re \sim 10^{3}$, smaller than the experiments by a factor of $10$ to $10^{3}$, because the computational effort scales with the cube of the target Reynolds number. The gap between shock-resolved viscous simulations and experiments remains to be bridged.

Inviscid simulations of triple shock reflection were performed under the same conditions, but without viscosity or heat conduction. The same resolution, number of refinement levels and refinement criteria were used, and they were marched to the same time as their viscous counterparts.

\subsection{Viscous simulation results}\label{sec:viscous results}

Plots of temperature are presented in figure \ref{fig:simulations:gamma=1.06,Mi,Re}. Each row represents one simulation at the Mach number specified in the first column. The Reynolds number is increased from $\approx 30$ to $\approx 110$ and $\approx 1000$ in each column. The Reynolds number is measured the same way as in experiments: equation \ref{eq:experiment:Reynolds number} is used with the shear velocity and viscosity from three shock theory and the Mach stem height is measured from the results.
The temperature scales on the right ranges from the incident shock temperature to the maximal temperature in the simulation. They apply to each sub-figure in the row. 
The sub-figures are cropped to the region of interest around the jet, vortex, and Mach stem.

\begin{figure}
	\centering
	\begin{subfigure}[t]{0.3\textwidth}
		\centering
		\includegraphics[scale=1]{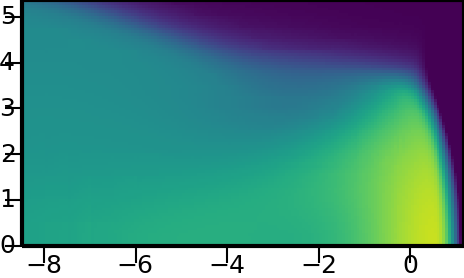}
		\caption{\label{fig:simulations:gamma=1.06,M=2.5,Re=10}$M_{\mathrm{i}} {=} 2.5$, $Re {=} 34$, $t{=}12$}
	\end{subfigure}%
	\begin{subfigure}[t]{0.3\textwidth}
		\centering
		\includegraphics[scale=1]{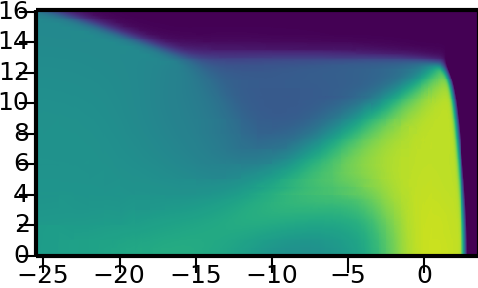}
		\caption{\label{fig:simulations:gamma=1.06,M=2.5,Re=100}$Re = 114$, $t=36$}
	\end{subfigure}%
	\begin{subfigure}[t]{0.3\textwidth}
		\centering
		\includegraphics[scale=1]{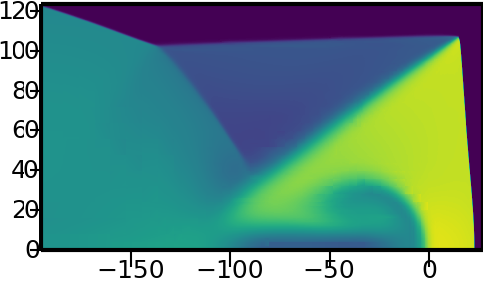}
		\caption{\label{fig:simulations:gamma=1.06,M=2.5,Re=1000}$Re = 1014$, $t=276$}
	\end{subfigure}%
	\begin{subfigure}[t]{0.1\textwidth}
		\centering
		\includegraphics[scale=1]{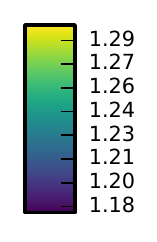}
	\end{subfigure}%
	\\
	\begin{subfigure}[t]{0.3\textwidth}
		\centering
		\includegraphics[scale=1]{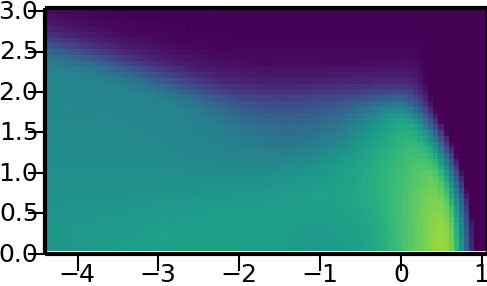}
		\caption{$M_{\mathrm{i}} {=} 3.0$, $Re {=} 29$, $t{=}6$}
	\end{subfigure}%
	\begin{subfigure}[t]{0.3\textwidth}
		\centering
		\includegraphics[scale=1]{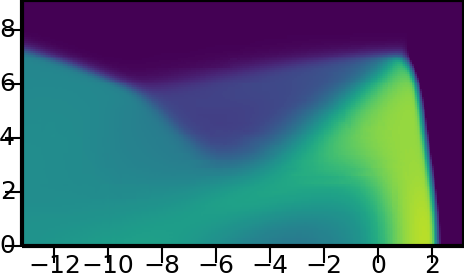}
		\caption{$Re = 108$, $t=18$}
	\end{subfigure}%
	\begin{subfigure}[t]{0.3\textwidth}
		\centering
		\includegraphics[scale=1]{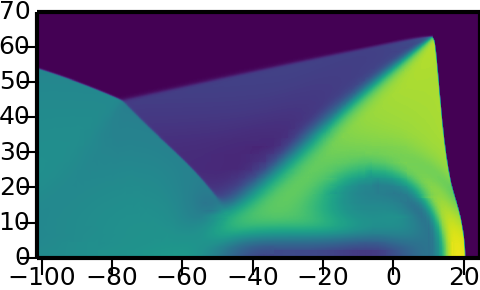}
		\caption{$Re = 993$, $t=138$}
	\end{subfigure}%
	\begin{subfigure}[t]{0.1\textwidth}
		\centering
		\includegraphics[scale=1]{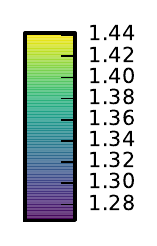}
	\end{subfigure}%
	\\
	\begin{subfigure}[t]{0.3\textwidth}
		\centering
		\includegraphics[scale=1]{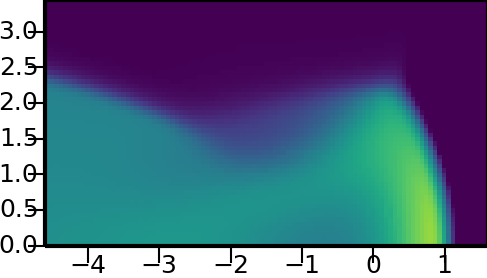}
		\caption{$M_{\mathrm{i}} {=} 3.5$, $Re {=} 48$, $t{=}6$}
	\end{subfigure}%
	\begin{subfigure}[t]{0.3\textwidth}
		\centering
		\includegraphics[scale=1]{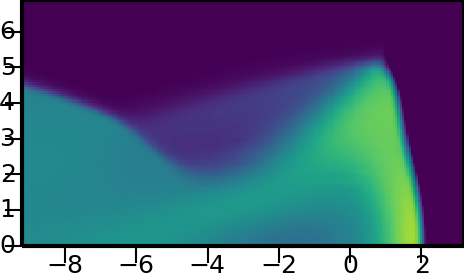}
		\caption{$Re = 118$, $t=12$}
	\end{subfigure}%
	\begin{subfigure}[t]{0.3\textwidth}
		\centering
		\includegraphics[scale=1]{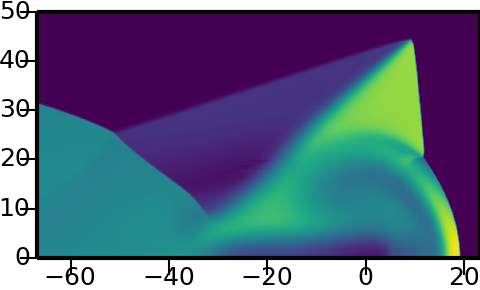}
		\caption{\label{fig:simulations:gamma=1.06,M=3.5,Re=1000}$Re = 1022$, $t=87$}
	\end{subfigure}%
	\begin{subfigure}[t]{0.1\textwidth}
		\centering
		\includegraphics[scale=1]{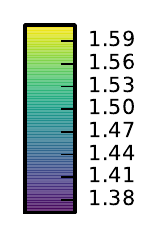}
	\end{subfigure}%
	\\
	\begin{subfigure}[t]{0.3\textwidth}
		\centering
		\includegraphics[scale=1]{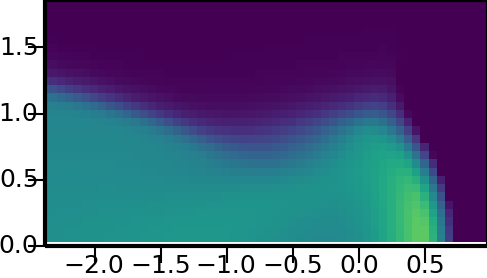}
		\caption{$M_{\mathrm{i}} {=} 4.0$, $Re {=} 32$, $t{=}3$}
	\end{subfigure}%
	\begin{subfigure}[t]{0.3\textwidth}
		\centering
		\includegraphics[scale=1]{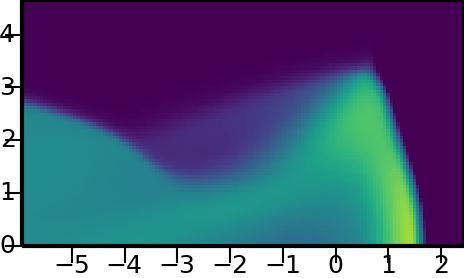}
		\caption{$Re = 103$, $t=7.5$}
	\end{subfigure}%
	\begin{subfigure}[t]{0.3\textwidth}
		\centering
		\includegraphics[scale=1]{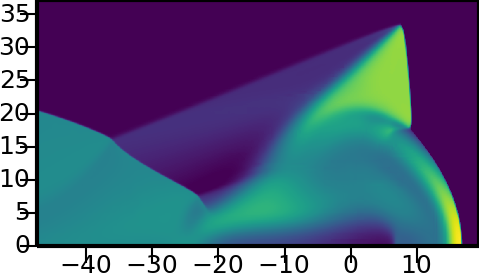}
		\caption{$Re = 1029$, $t=60$}
	\end{subfigure}%
	\begin{subfigure}[t]{0.1\textwidth}
		\centering
		\includegraphics[scale=1]{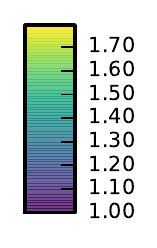}
	\end{subfigure}%
	\\
	\begin{subfigure}[t]{0.3\textwidth}
		\centering
		\includegraphics[scale=1]{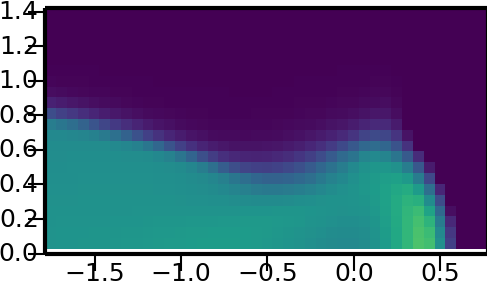}
		\caption{$M_{\mathrm{i}} {=} 5.0$, $Re {=} 31$, $t{=}2$}
	\end{subfigure}%
	\begin{subfigure}[t]{0.3\textwidth}
		\centering
		\includegraphics[scale=1]{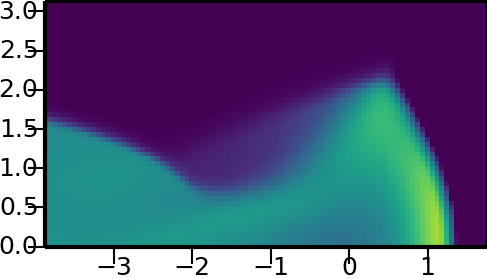}
		\caption{$Re = 101$, $t=4.5$}
	\end{subfigure}%
	\begin{subfigure}[t]{0.3\textwidth}
		\centering
		\includegraphics[scale=1]{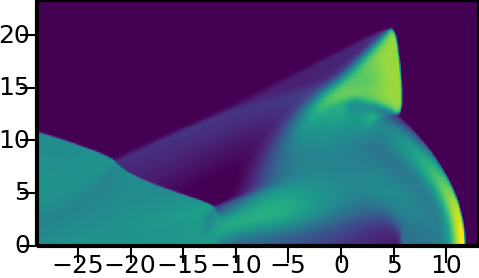}
		\caption{$Re = 965$, $t=33$}
	\end{subfigure}%
	\begin{subfigure}[t]{0.1\textwidth}
		\centering
		\includegraphics[scale=1]{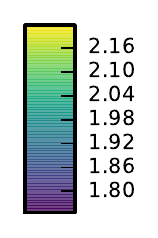}
	\end{subfigure}%
	\\
	\begin{subfigure}[t]{0.3\textwidth}
		\centering
		\includegraphics[scale=1]{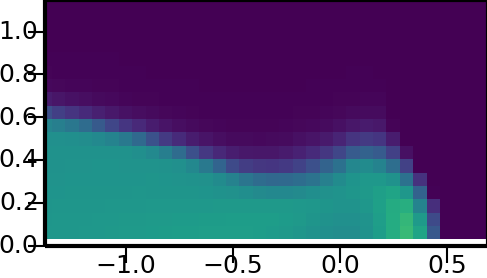}
		\caption{$M_{\mathrm{i}} {=} 6.0$, $Re {=} 29$, $t{=}1.5$}
	\end{subfigure}%
	\begin{subfigure}[t]{0.3\textwidth}
		\centering
		\includegraphics[scale=1]{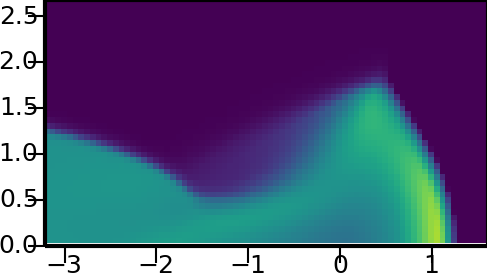}
		\caption{$Re = 106$, $t=3.5$}
	\end{subfigure}%
	\begin{subfigure}[t]{0.3\textwidth}
		\centering
		\includegraphics[scale=1]{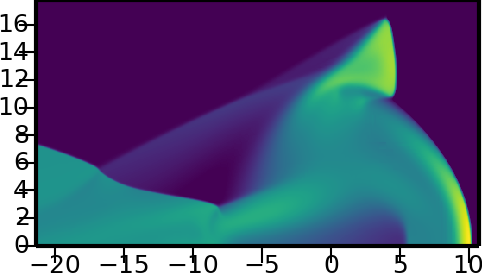}
		\caption{\label{fig:simulations:gamma=1.06,M=6.0,Re=1000}$Re = 1010$, $t=24$}
	\end{subfigure}%
	\begin{subfigure}[t]{0.1\textwidth}
		\centering
		\includegraphics[scale=1]{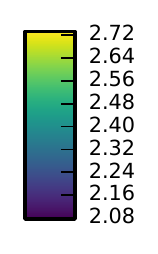}
	\end{subfigure}%
	\caption{\label{fig:simulations:gamma=1.06,Mi,Re}Temperature plots of viscous simulations ($\gamma=1.06$, $\theta_{\mathrm{w}} = 30^{\circ}$)}
\end{figure}

The triple shock reflection with an incident shock of $M_{\mathrm{i}}=2.5$ is shown in figures \ref{fig:simulations:gamma=1.06,M=2.5,Re=10} to \ref{fig:simulations:gamma=1.06,M=2.5,Re=1000}. The Mach stem at $Re = 34$ is smoothly curved from the triple point to the reflecting surface and straightens as Reynolds number increases. The forward jet grows with Reynolds number and develops a vortex by $Re \approx 1000$, however, the head of the forward jet remains far from the Mach stem.

The jet is closer to the Mach stem at $M_{\mathrm{i}}=3.0$. The jet reaches the Mach stem and causes it to bulge when $Re \approx 1000$, and eventually bifurcate when $Re > 1200$. The vortex at the head of the jet is larger than the previous case.

At $M_{\mathrm{i}}=3.5$, the forward jet is strong enough to cause a change of curvature in the Mach stem by $Re \approx 110$, and the Mach stem bifurcates when $Re > 440$, as seen in figure \ref{fig:simulations:gamma=1.06,M=3.5,Re=1000}. Only a portion of the Mach stem is straight below the triple point. A new triple point is located about midway on the Mach stem, below which the Mach stem is curved.
The vortex is large enough that it causes the slip line to deflect downwards.
Double Mach reflections with a Mach stem bifurcation have been classified as triple Mach-White reflections by \cite{semenov_classification_2009a}.

The trend continues as Mach number is increased: the jet moves closer to the Mach stem, causing the Mach stem to deform and bifurcate as early as $Re \approx 200$. The jet terminates in an increasingly large vortex that dominates the space behind the Mach stem. The vortex grows large enough to interfere with the slip line and flow behind the reflected shock.

\subsection{Inviscid simulation results}\label{sec:inviscid results}

The Euler equations are typically used to simulate shock reflections and detonations. These `inviscid' simulations suffer from numerical dissipation that depends on grid and scheme, not a physical phenomenon. However, inviscid simulations are much faster to compute and offer insight on how the reflection will evolve as Reynolds number becomes very large.

\begin{figure}
	\centering
	\begin{subfigure}[t]{0.3\textwidth}
		\centering
		\includegraphics[scale=1]{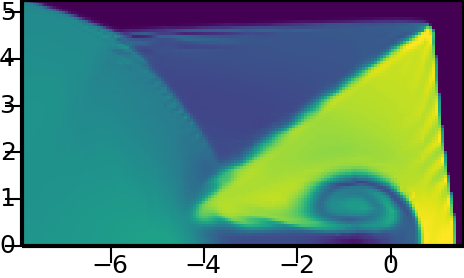}
		\caption{$M_{\mathrm{i}} = 2.5$, $t=12$}
	\end{subfigure}%
	\begin{subfigure}[t]{0.3\textwidth}
		\centering
		\includegraphics[scale=1]{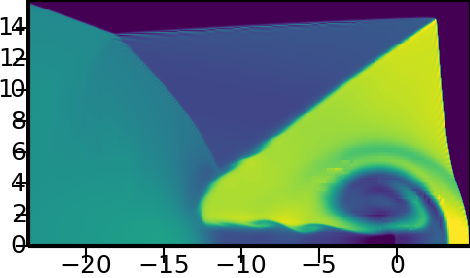}
		\caption{$t=36$}
	\end{subfigure}%
	\begin{subfigure}[t]{0.3\textwidth}
		\centering
		\includegraphics[scale=1]{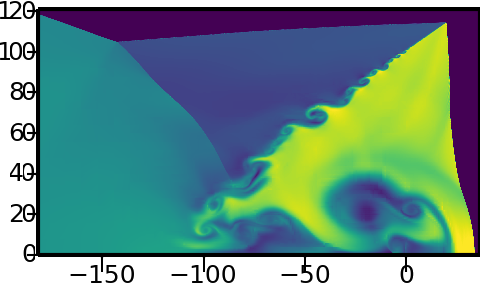}
		\caption{$t=276$}
	\end{subfigure}%
	\begin{subfigure}[t]{0.1\textwidth}
		\centering
		\includegraphics[scale=1]{{./figures/simulations/g=1.06,w0=2.61799387833333333333,M=2.5,Pr=0.75/T_colorbar_norm}.pdf}
	\end{subfigure}%
	\\
	\begin{subfigure}[t]{0.3\textwidth}
		\centering
		\includegraphics[scale=1]{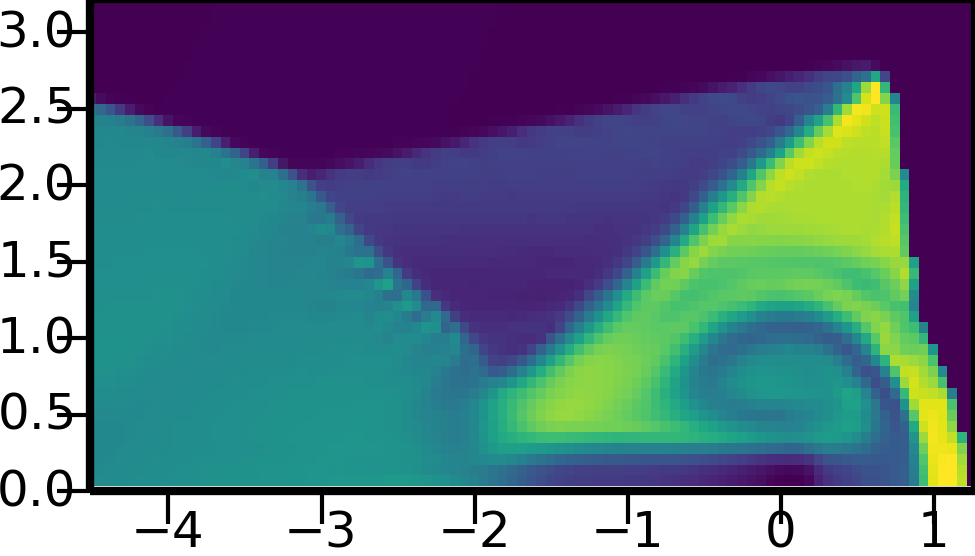}
		\caption{$M_{\mathrm{i}} = 3.0$, $t=6$}
	\end{subfigure}%
	\begin{subfigure}[t]{0.3\textwidth}
		\centering
		\includegraphics[scale=1]{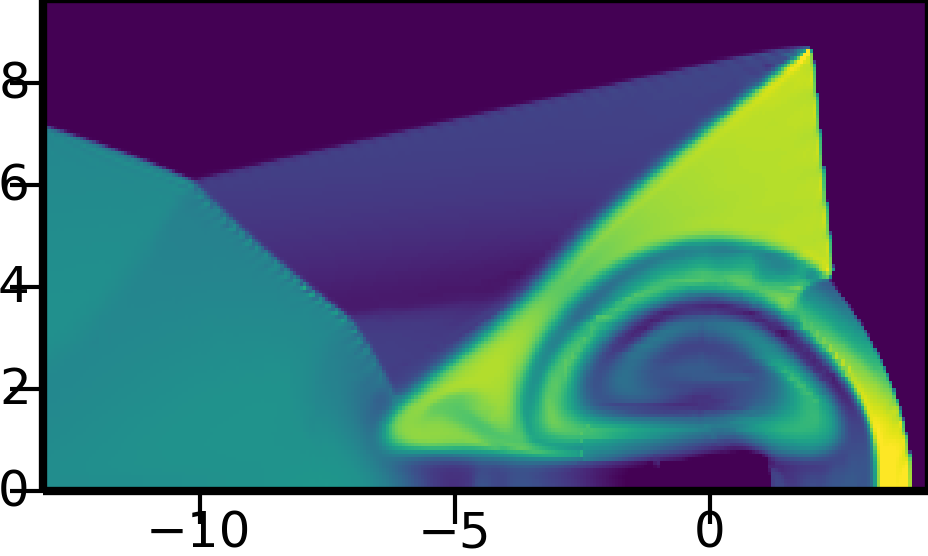}
		\caption{$t=18$}
	\end{subfigure}%
	\begin{subfigure}[t]{0.3\textwidth}
		\centering
		\includegraphics[scale=1]{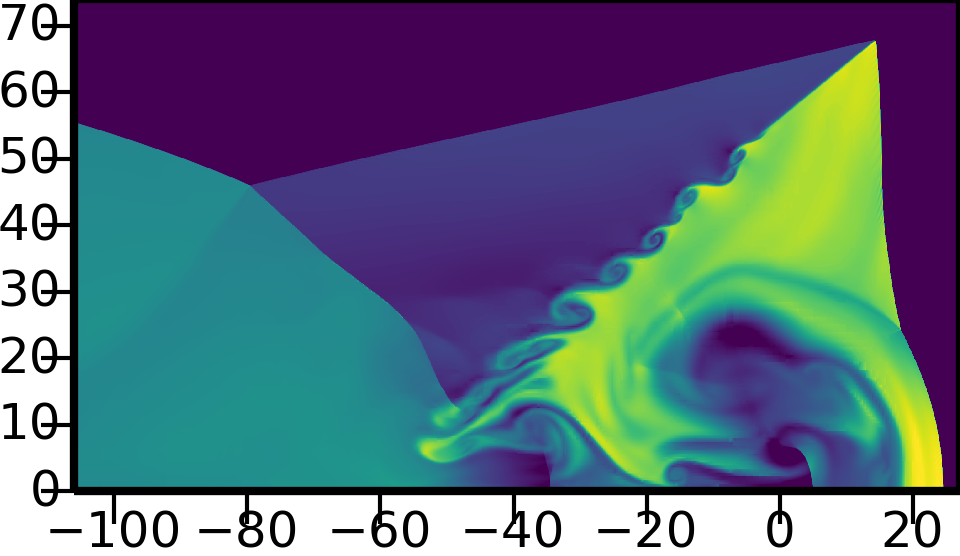}
		\caption{$t=138$}
	\end{subfigure}%
	\begin{subfigure}[t]{0.1\textwidth}
		\centering
		\includegraphics[scale=1]{{./figures/simulations/g=1.06,w0=2.61799387833333333333,M=3.0,Pr=0.75/T_colorbar_norm}.pdf}
	\end{subfigure}%
	\\
	\begin{subfigure}[t]{0.3\textwidth}
		\centering
		\includegraphics[scale=1]{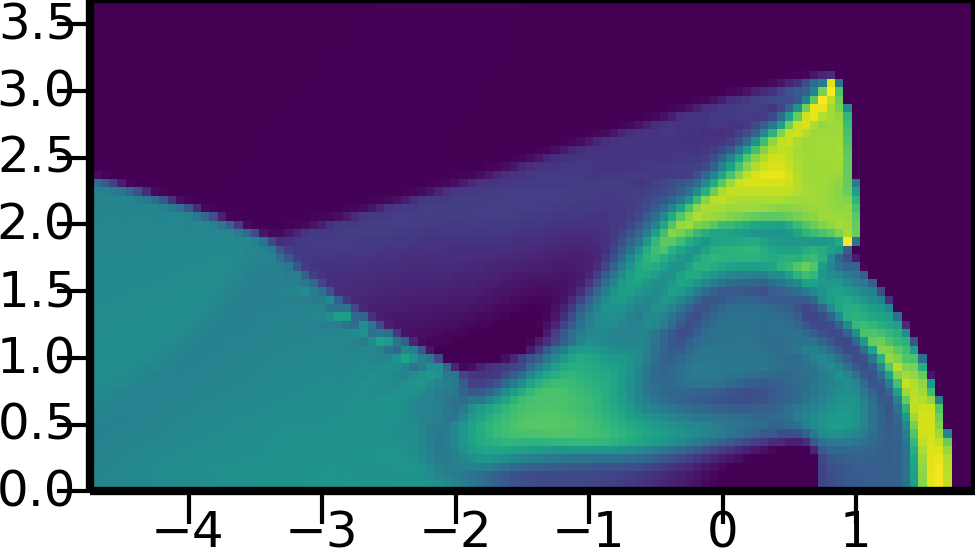}
		\caption{$M_{\mathrm{i}} = 3.5$, $t=6$}
	\end{subfigure}%
	\begin{subfigure}[t]{0.3\textwidth}
		\centering
		\includegraphics[scale=1]{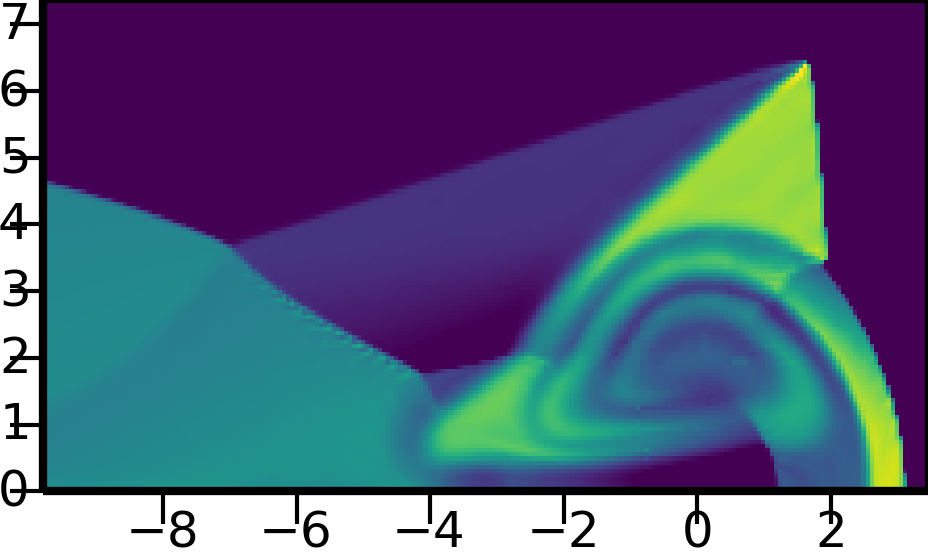}
		\caption{$t=12$}
	\end{subfigure}%
	\begin{subfigure}[t]{0.3\textwidth}
		\centering
		\includegraphics[scale=1]{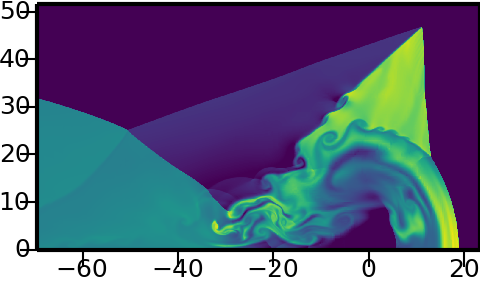}
		\caption{$t=84$}
	\end{subfigure}%
	\begin{subfigure}[t]{0.1\textwidth}
		\centering
		\includegraphics[scale=1]{{./figures/simulations/g=1.06,w0=2.61799387833333333333,M=3.5,Pr=0.75/T_colorbar_norm}.pdf}
	\end{subfigure}
	\\
	\begin{subfigure}[t]{0.3\textwidth}
		\centering
		\includegraphics[scale=1]{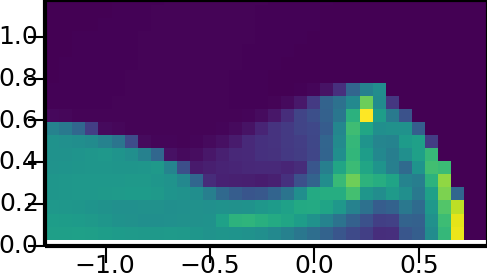}
		\caption{\label{fig:simulations:gamma=1.06,Mi=6,inviscid,t=1.5}$M_{\mathrm{i}} = 6$, $t=1.5$}
	\end{subfigure}%
	\begin{subfigure}[t]{0.3\textwidth}
		\centering
		\includegraphics[scale=1]{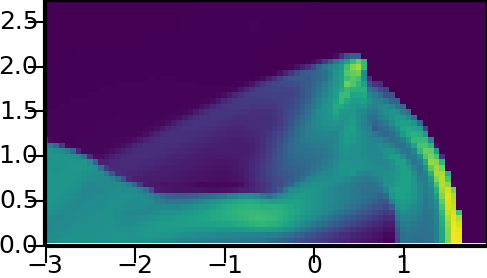}
		\caption{$t=3.5$}
	\end{subfigure}%
	\begin{subfigure}[t]{0.3\textwidth}
		\centering
		\includegraphics[scale=1]{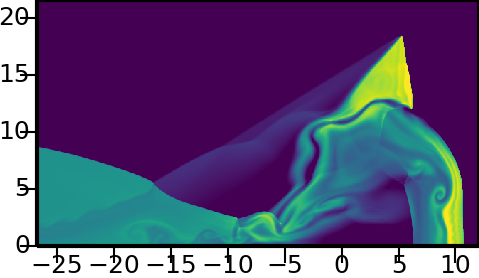}
		\caption{\label{fig:simulations:gamma=1.06,Mi=6,inviscid,t=24}$t=24$}
	\end{subfigure}%
	\begin{subfigure}[t]{0.1\textwidth}
		\centering
		\includegraphics[scale=1]{{./figures/simulations/g=1.06,w0=2.61799387833333333333,M=6.0,Pr=0.75/T_colorbar_norm}.pdf}
	\end{subfigure}%
	\caption{\label{fig:simulations:gamma=1.06,Mi,inviscid}Temperature plots of inviscid simulations ($\gamma=1.06$, $\theta_{\mathrm{w}} = 30^{\circ}$)}
\end{figure}

Temperature plots of inviscid simulations are shown in figure \ref{fig:simulations:gamma=1.06,Mi,inviscid}. The first row shows results for $M_{\mathrm{i}} = 2.5$. The inviscid double Mach reflection at $t = 12$ resembles the viscous case in figure \ref{fig:simulations:gamma=1.06,M=2.5,Re=1000}, but with Kelvin-Helmholtz instabilities along the slip line and a vortex that has completed more than one rotation. As time increases, the forward jet and vortex move closer to the Mach stem, causing a change of curvature on the Mach stem but no bifurcation. Kelvin-Helmholtz instability grow along the slip line, forward jet, and vortex.

The following rows of figure \ref{fig:simulations:gamma=1.06,Mi,inviscid} show the effect of increasing Mach number. The jet approaches, deforms, and bifurcates the Mach stem; the vortex becomes large enough to disrupt the slip line, and a shock develops in the forward jet. Kelvin-Helmholtz instabilities become more prevalent in the vortex, resulting in a heterogeneous temperature field behind the Mach stem. 

At larger Mach numbers ($M_{\mathrm{i}} \ge 4.5$) the bifurcation point is disproportionately elevated at early times (\textit{e.g.} figures \ref{fig:simulations:gamma=1.06,Mi=6,inviscid,t=1.5} to \ref{fig:simulations:gamma=1.06,Mi=6,inviscid,t=24}). 
Here the bifurcation point moves up-and-down and the Mach stem foot rocks to-and-fro, leading to instances where the Mach stem bulge appears flattened, like in figure \ref{fig:simulations:gamma=1.06,Mi=6,inviscid,t=24}. However, the shape of the bifurcated Mach stem generally remains similar to figure \ref{fig:simulations:gamma=1.06,M=6.0,Re=1000}, composed of a straight Mach shock and round bulge.

\subsection{Summary of simulations}

The reflection of a triple point from an axis of symmetry was simulated for $M_{\mathrm{i}} = 2.5$ to $6$, $\gamma = 1.06$, $\theta_{\mathrm{w}} = 30^{\circ}$, and $Re \le 2\EXP{3}$. 
Reynolds number was found to play and important role in the development of the shock reflection. Forward jetting and the vortex size increased with Reynolds number and Mach number. Interaction of the jet with the Mach stem led to bulging and bifurcation of the Mach stem. Inviscid simulations developed Kelvin-Helmholtz instability and heterogeneous temperature fields in the vortex behind the Mach stem that were absent from the viscous results. Varying the Mach number led to the same changes that were seen in experiments.

\section{Discussion}\label{sec:discussion}

The experiments and numerics are compared in the next two subsections. A qualitative comparison of the shock structures is made in section \ref{sec:Comparison of simulations}, followed by a comparison of bulging and jetting, and of numerical triple point paths. Mach stem bifurcation is compared in section \ref{sec:Mach stem bifurcation limits} then explored in further depth using inviscid simulations.

The reader should remain cognizant of differences between viscous simulations where $Re \lesssim 10^3$, inviscid simulations that hint at larger Reynolds numbers, and unsteady experiments where $Re \sim 10^5$. The viscous simulations resemble triple shock reflections that occur at the beginning of a detonation cell, and the experiments, which include shock curvature, are a better representation of the reflection process later in the detonation cell cycle. Differences between individual simulations and experiments are to be expected. However, the effects that lead to a strong forward jet, vortex, and jet-shock interaction can be inferred by comparing the three sets of results while considering their similarities, their trends, and the unique ways they differ.

\subsection{\label{sec:Comparison of simulations}Comparison of results}

\begin{figure}
	\centering
	\begin{subfigure}[t]{0.33\textwidth}
		\centering
		Viscous simulation
	\end{subfigure}%
	\begin{subfigure}[t]{0.33\textwidth}
		\centering
		Inviscid simulation
	\end{subfigure}%
	\begin{subfigure}[t]{0.33\textwidth}
		\centering
		Experiment
	\end{subfigure}
	\\
	\begin{subfigure}[t]{0.33\textwidth}
		\centering
		\includegraphics[scale=0.37]{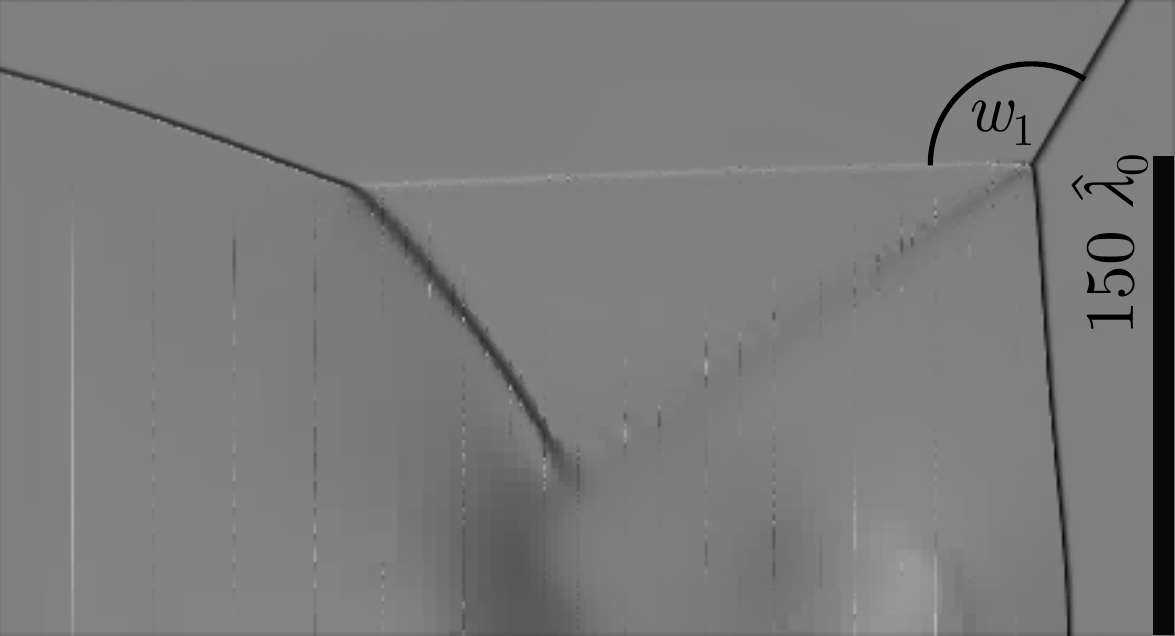}
		\caption{$M_{\mathrm{i}}=2.5$, $Re = 1340$}
	\end{subfigure}%
	\begin{subfigure}[t]{0.33\textwidth}
		\centering
		\includegraphics[scale=0.37]{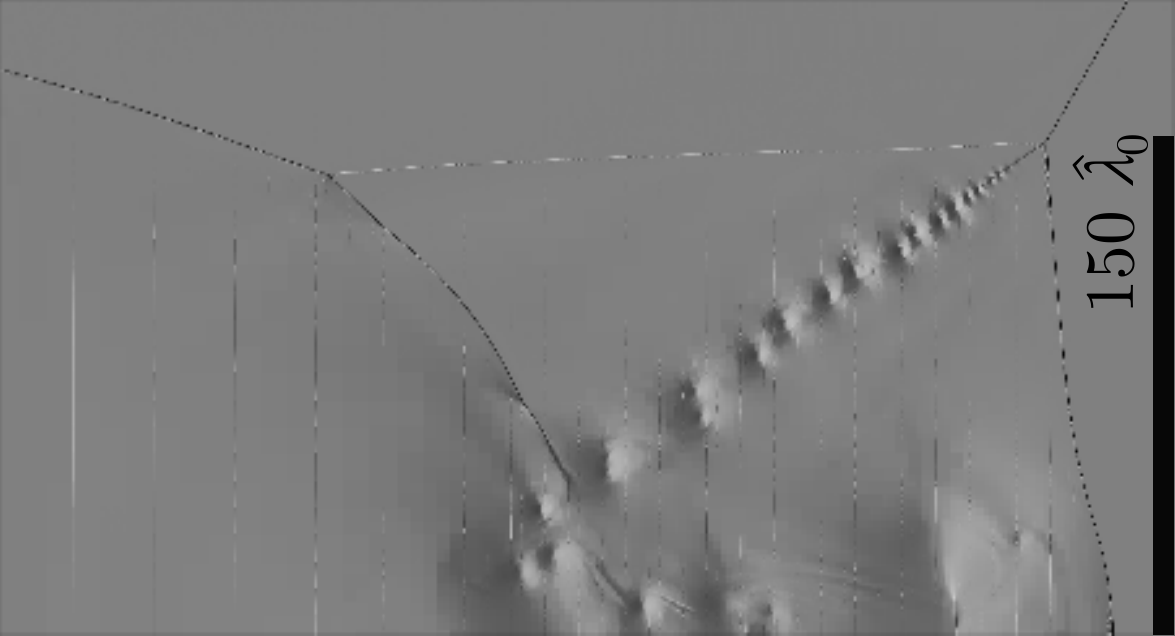}
		\caption{$M_{\mathrm{i}}=2.5$}
	\end{subfigure}%
	\begin{subfigure}[t]{0.33\textwidth}
		\centering
		\includegraphics[scale=1.71]{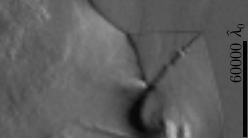}
		\caption{$M_{\mathrm{c}}=2.5$, $Re = 6.8 \EXP{5}$}
	\end{subfigure}%
	\\
	\begin{subfigure}[t]{0.33\textwidth}
		\centering
		\includegraphics[scale=0.37]{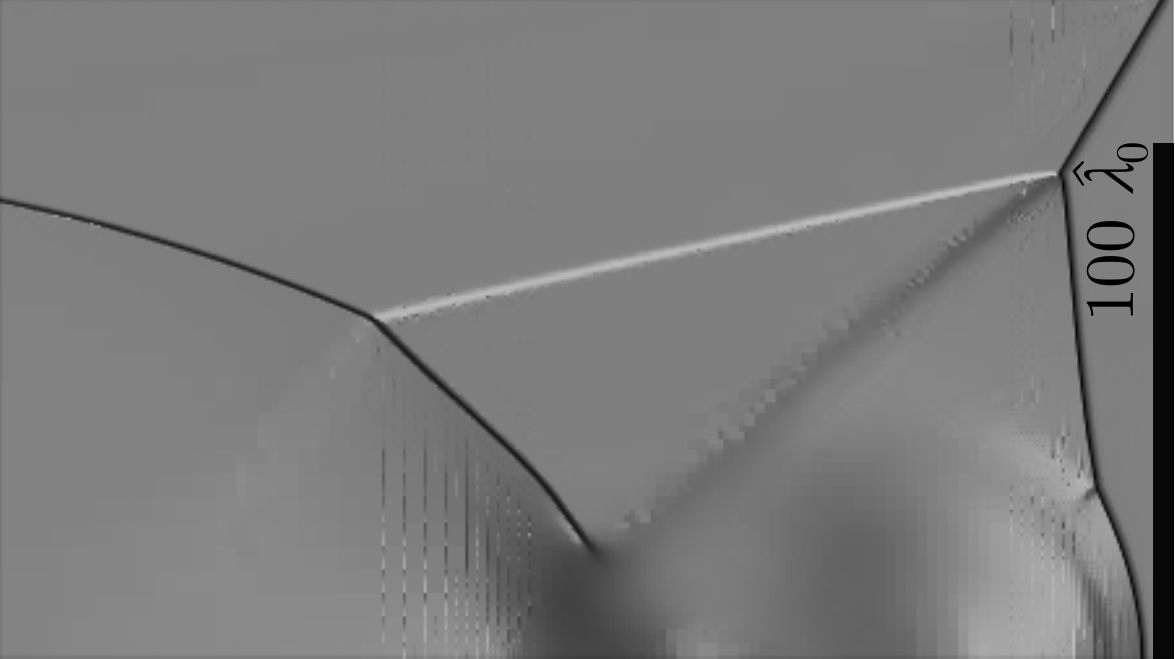}
		\caption{$M_{\mathrm{i}}=3.0$, $Re = 1481$}
	\end{subfigure}%
	\begin{subfigure}[t]{0.33\textwidth}
		\centering
		\includegraphics[scale=0.37]{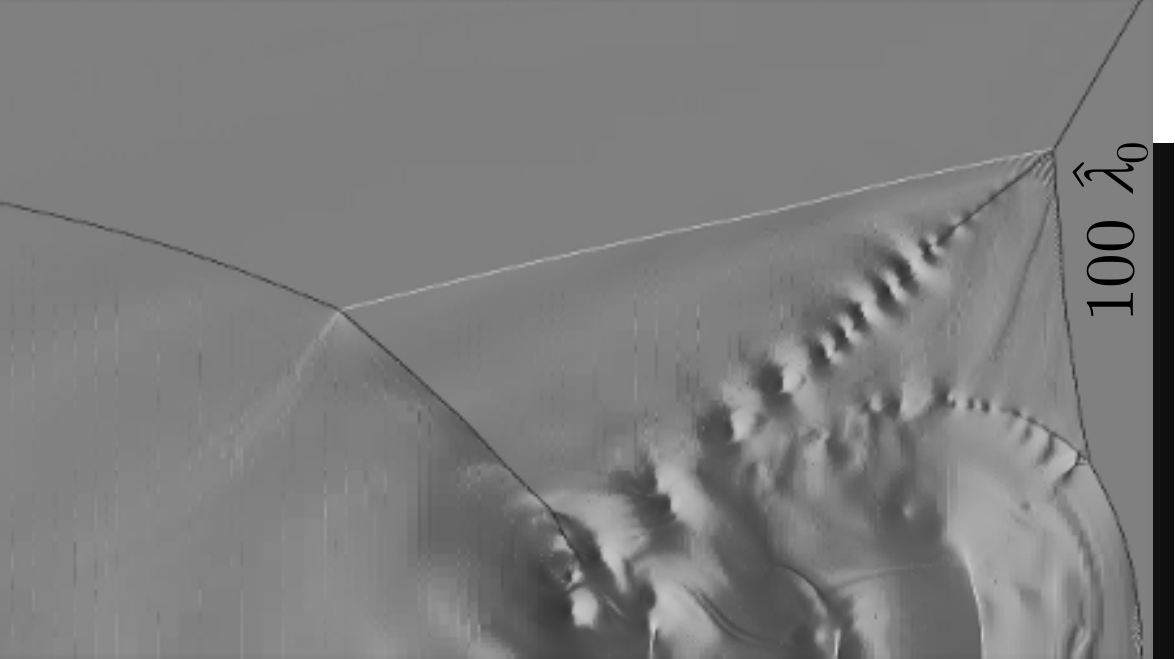}
		\caption{$M_{\mathrm{i}}=3.0$}
	\end{subfigure}%
	\begin{subfigure}[t]{0.33\textwidth}
		\centering
		\includegraphics[scale=2.275]{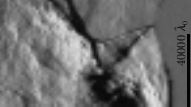}
		\caption{$M_{\mathrm{c}}=2.7$, $Re = 5.2 \EXP{5}$}
	\end{subfigure}%
	\\
	\begin{subfigure}[t]{0.33\textwidth}
		\centering
		\includegraphics[scale=0.39]{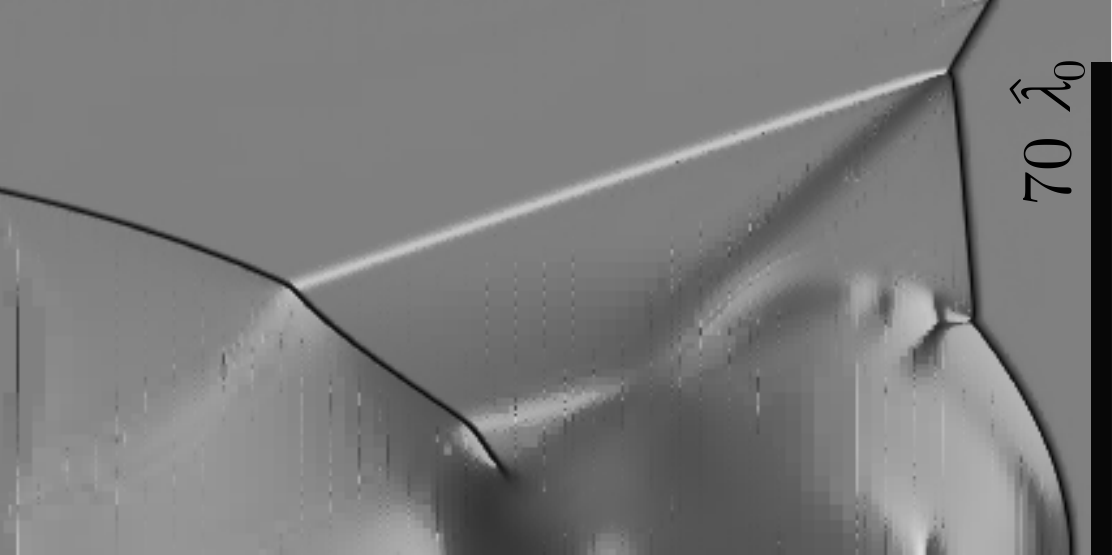}
		\caption{$M_{\mathrm{i}}=3.5$, $Re = 1600$}
	\end{subfigure}%
	\begin{subfigure}[t]{0.33\textwidth}
		\centering
		\includegraphics[scale=0.3873]{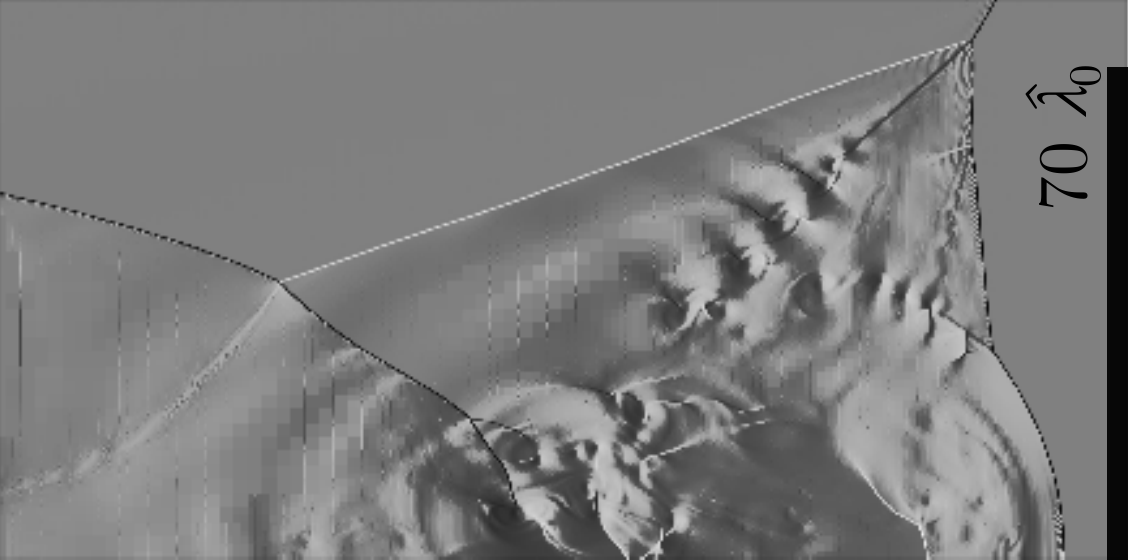}
		\caption{$M_{\mathrm{i}}=3.5$}
	\end{subfigure}%
	\begin{subfigure}[t]{0.33\textwidth}
		\centering
		\includegraphics[scale=2.32]{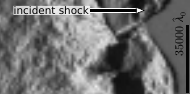}
		\caption{$M_{\mathrm{c}}=3.4$, $Re = 7.5 \EXP{5}$}
	\end{subfigure}%
	\caption{\label{fig:comparison}Comparison of viscous (left column) and inviscid (centre column) simulations with experiments (right column) ($\gamma = 1.06$, $\theta_{\mathrm{w}} = 30^{\circ}$); scale shown by vertical black bar}
\end{figure}

Experiments and simulations are qualitatively compared in figure \ref{fig:comparison}. As predicted by three shock theory and observed in all cases, increasing Mach number shortens the Mach stem and increases the reflected shock angle $w_1$. More pertinent to this study, raising the Mach number also increases the forward jet, vortex, and bulging size in all cases. The forward jet catches up to the shock front ($M \ge 2.7$), causing it to deform in the experiment ($M_{\mathrm{c}} = 3.4$) and clearly bifurcate in simulations.

The forward jet length, size of the vortex, and the shock reflection structure are qualitatively alike in the inviscid simulations and the viscous simulations at $Re > 1000$. However, the inviscid simulations contain Kelvin-Helmholtz instability that significantly change the flow field in the jet and vortex behind the Mach stem. The instabilities cause rough-looking features in the flow field behind Mach stem in experiments and inviscid simulations, but are absent from viscous simulations. This means Reynolds number plays an important role in the development of large-scale mixing behind the Mach stem.

\begin{figure}
	\centering
	\begin{subfigure}[t]{0.65\textwidth}
		\centering
		\includegraphics[scale=1]{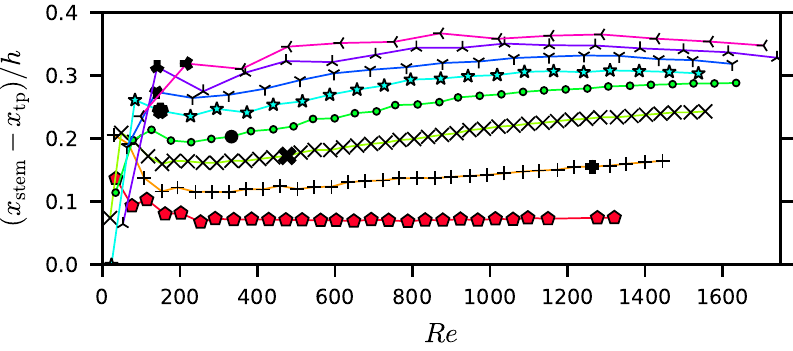}
		\caption{\label{fig:simulation:forward_jetting_quantitative:bulging}Mach stem position; left: viscous, right: inviscid}
	\end{subfigure}%
	\begin{subfigure}[t]{0.2\textwidth}
		\includegraphics[scale=1]{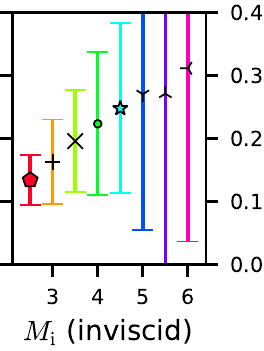}
	\end{subfigure}
	\\\vspace{0.25cm}
	\begin{subfigure}[t]{0.65\textwidth}
		\centering
		\includegraphics[scale=1]{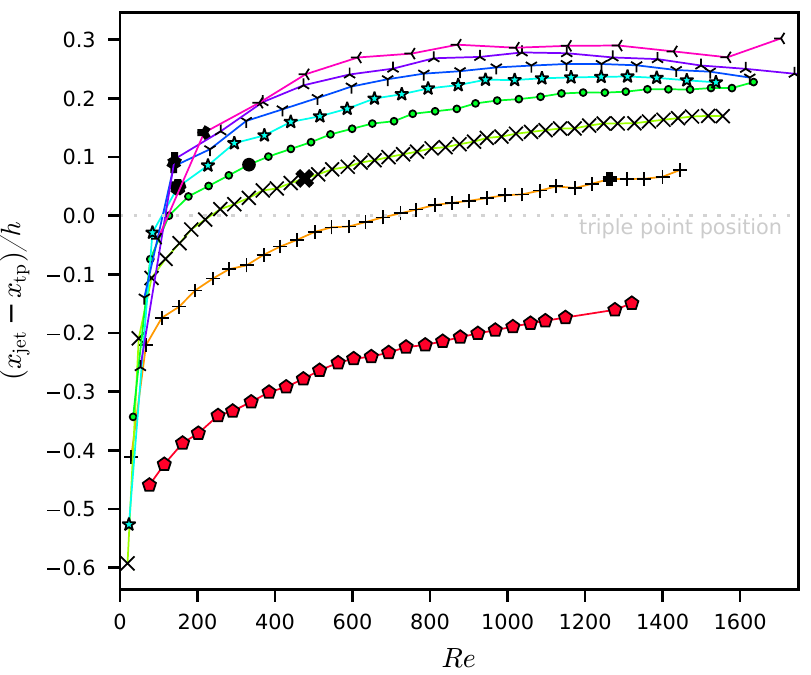}
		\caption{Viscous jet position}
	\end{subfigure}%
	\begin{subfigure}[t]{0.2\textwidth}
		\centering
		\vspace{-5cm}
		\includegraphics[scale=1]{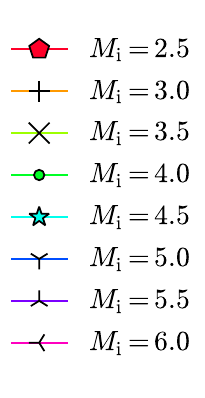}
	\end{subfigure}
	\caption{\label{fig:simulation:forward_jetting_quantitative}Evolution of the Mach stem and jet at the reflecting surface from simulations ($\gamma=1.06$, $\theta_{\mathrm{w}} = 30^{\circ}$); bold points where bifurcation is first observed; error bars at 3 standard deviations}
\end{figure}

The extent of forward jetting and Mach stem bulging in simulations are quantified in figure \ref{fig:simulation:forward_jetting_quantitative} as functions of the Reynolds and Mach numbers. They are measured as the horizontal distance between the triple point and the jet's head, or Mach stem position along the reflecting surface, respectively, and normalized by the Mach stem height. The amount of viscous bulging on the left of figure \ref{fig:simulation:forward_jetting_quantitative:bulging} is compared to the average inviscid value on the right, sharing the ordinate axis. Figure \ref{fig:simulation:forward_jetting_quantitative} shows that the viscous Mach stem and jet overtake the triple point as Mach number and Reynolds number are increased. The jet size and amount of bulging grow in tandem once the jet passes the triple point. The amount of bulging converges towards the mean inviscid value.

The Mach stem bulges more in simulations than experiments (figure \ref{fig:experiment:quantitative:g=1.06}). 
While some difference is to be expected due to the experiments' unsteadiness and large Reynolds numbers, the difference remains to be reconciled.
It is worth noting three-dimensional effects caused by the shallow channel depth (19.1 mm) are not responsible for this difference. In experiment 9, for example, while the shock travels 200 mm from the tip of the obstacle, boundary layers on the channel windows grow only 0.7 mm (\cite{fay1959two}) from the shock to the rear of the jet. The boundary layers do not intersect across the jet.

\begin{figure}
	\centering
	\includegraphics[scale=0.8]{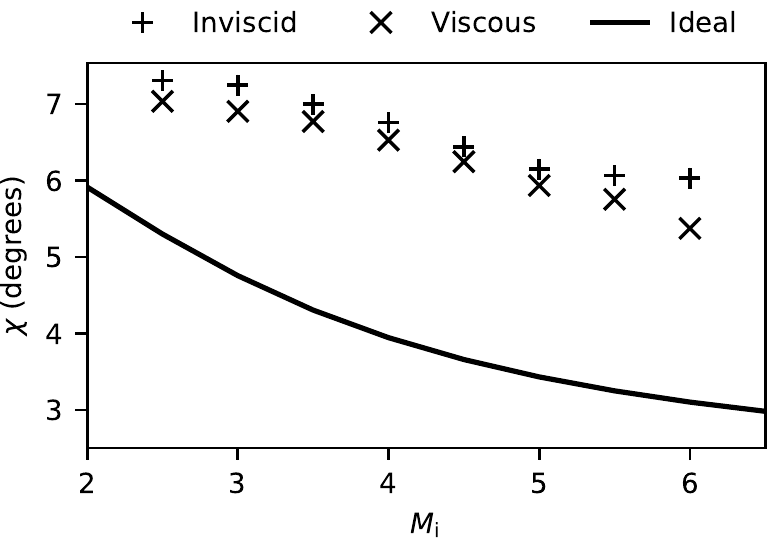}
	\caption{\label{fig:simulation:triple point path}Angle of the triple point path with respect to the reflecting surface from three shock theory, viscous and inviscid simulations ($\gamma=1.06$, $\theta_{\mathrm{w}} = 30^{\circ}$)}
\end{figure}

Another clear and easily-quantifiable measure is the median instantaneous angle $\chi$ between the triple point path and the horizontal plotted in figure \ref{fig:simulation:triple point path}. Experimental results are omitted due to their unsteadiness.  
The difference between viscous and inviscid simulations is less than one degree, suggesting viscosity has little effect on the triple point path. {\color{black}Three shock theory underestimates $\chi$ in the simulations by $1^{\circ}$ to $3^{\circ}$, increasing with Mach number, which is on par with its underestimation of experiments (\cite{andoPseudoStationaryObliqueShockWave1981}, $\gamma = 1.29$). This may be explained by the idealistic assumption that the Mach stem is perpendicular to the reflecting surface (\cite{li_analysis_1999}).}

\subsection{\label{sec:Mach stem bifurcation limits}Mach stem bifurcation limits}

Triple Mach-White reflections (\cite{semenov_classification_2009a}) are characterized by the appearance of a new triple point that bifurcates the Mach stem, however, there is little data (\cite{mach_bifurcating_2011}) on its limits. The limits of Mach stem bifurcations are explored in this section as a function of $M$, $\theta_{\mathrm{w}}$, $\gamma$ and $Re$.

\begin{figure}
	\centering
	\includegraphics[scale=1]{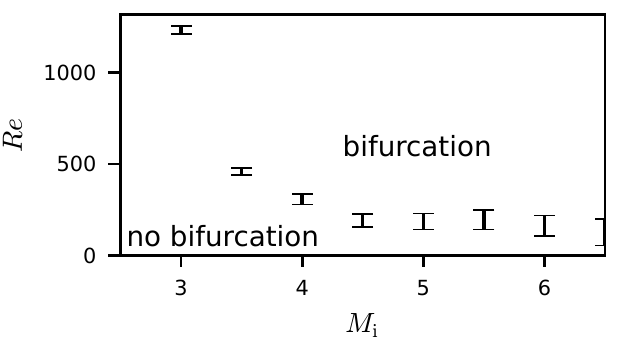}
	\caption{\label{fig:simulation:bifurcation_onset:Re}Onset of Mach stem bifurcation in viscous simulations ($\gamma=1.06$, $\theta_{\mathrm{w}} = 30^{\circ}$)}
\end{figure}

The Reynolds number where Mach stem bifurcation is first observed is displayed in figure \ref{fig:simulation:bifurcation_onset:Re}. Bifurcation occurs at Reynolds numbers above the points in the plot, while the Mach stem remains unbifurcated below. The plot points to a minimum Mach number, $M_{\mathrm{i}} < 3$, below which bifurcation does not occur when $\gamma = 1.06$ and $\theta_{\mathrm{w}} = 30^{\circ}$. These novel results show that Mach stem bifurcation is sensitive to Reynolds number near the minimum Mach number, whereas bifurcations at high Mach numbers occur early in the reflection, once $Re \gtrsim 200$. These points are plotted in bold in figure \ref{fig:simulation:forward_jetting_quantitative}, revealing that bifurcations occur when the jet overtakes the triple point by $(x_{\mathrm{jet}} - x_{\mathrm{tp}})/h \gtrsim 5\%$ (for $\gamma=1.06$, $\theta_{\mathrm{w}} = 30^{\circ}$). The presence of Mach stem bifurcation serves as an indication of the jet's strength.

There is agreement between inviscid simulations and viscous simulations  at $Re \gtrsim 1000$ regarding the shape of the shock reflection, the triple point path angle, and the amount of Mach stem bulging. Additionally, Mach stem bifurcation occurs between $2.5 \le M_{\mathrm{i}} < 3$ in both cases. This motivates the use of inviscid simulations to estimate the bifurcation limits, and the amount of jetting by proxy, over a range of reflection angles and isentropic exponents that would be computationally expensive to calculate with viscosity.

\begin{figure}
	\centering
	\begin{subfigure}[t]{0.33\textwidth}
		\centering
		\includegraphics[scale=1]{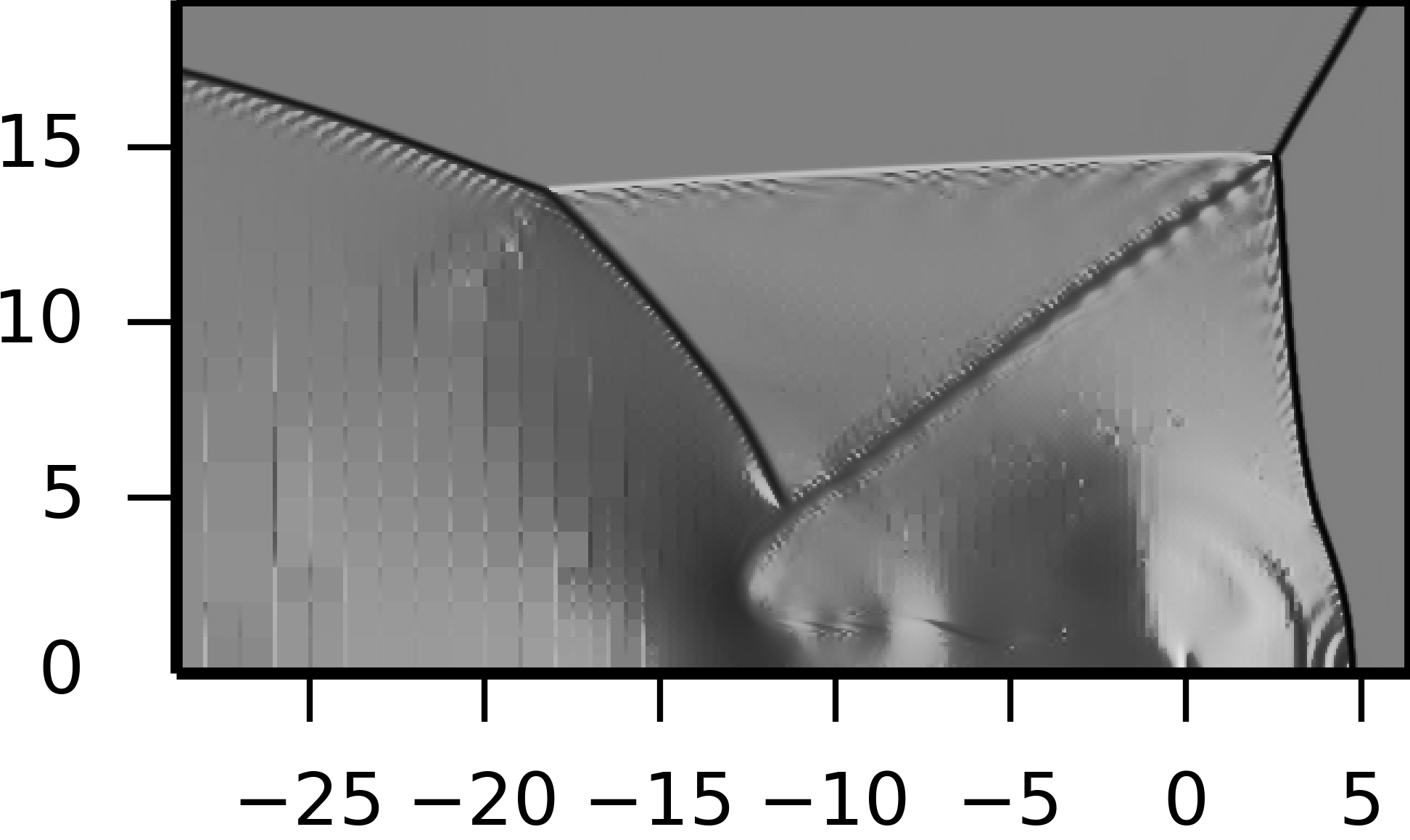}
		\caption{$M_{\mathrm{i}} = 2.5$, $t=36.3$}
	\end{subfigure}%
	\begin{subfigure}[t]{0.33\textwidth}
		\centering
		\includegraphics[scale=1.]{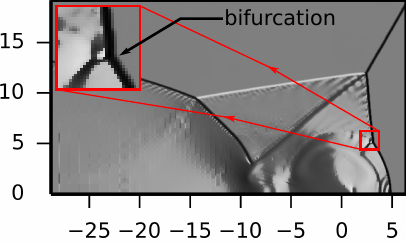}
		\caption{$M_{\mathrm{i}} = 2.75$, $t=26.5$}
	\end{subfigure}%
	\begin{subfigure}[t]{0.33\textwidth}
		\centering
		\includegraphics[scale=1.]{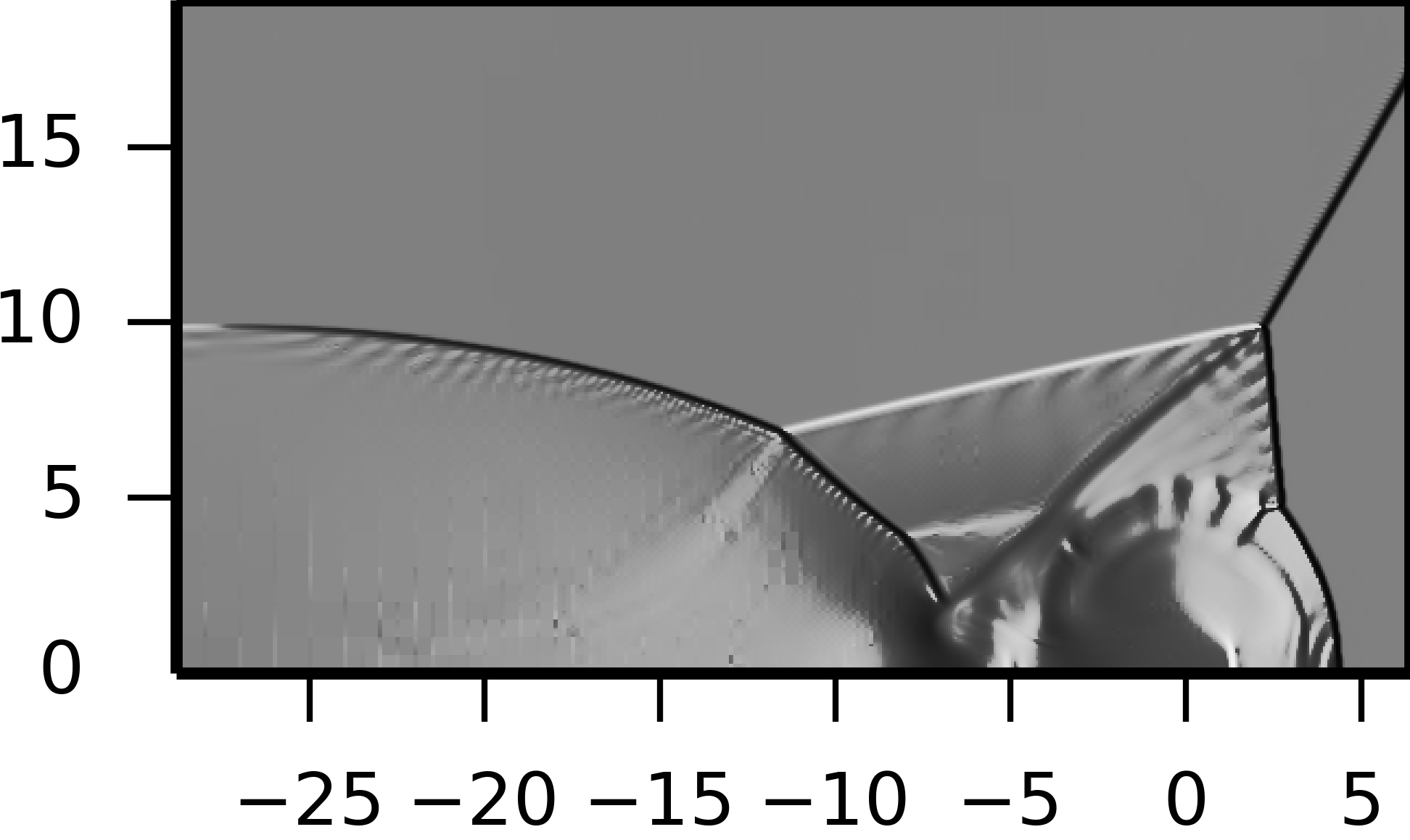}
		\caption{$M_{\mathrm{i}} = 3$, $t=20.3$}
	\end{subfigure}%
\\
	\begin{subfigure}[t]{0.33\textwidth}
		\centering
		\includegraphics[scale=1.02]{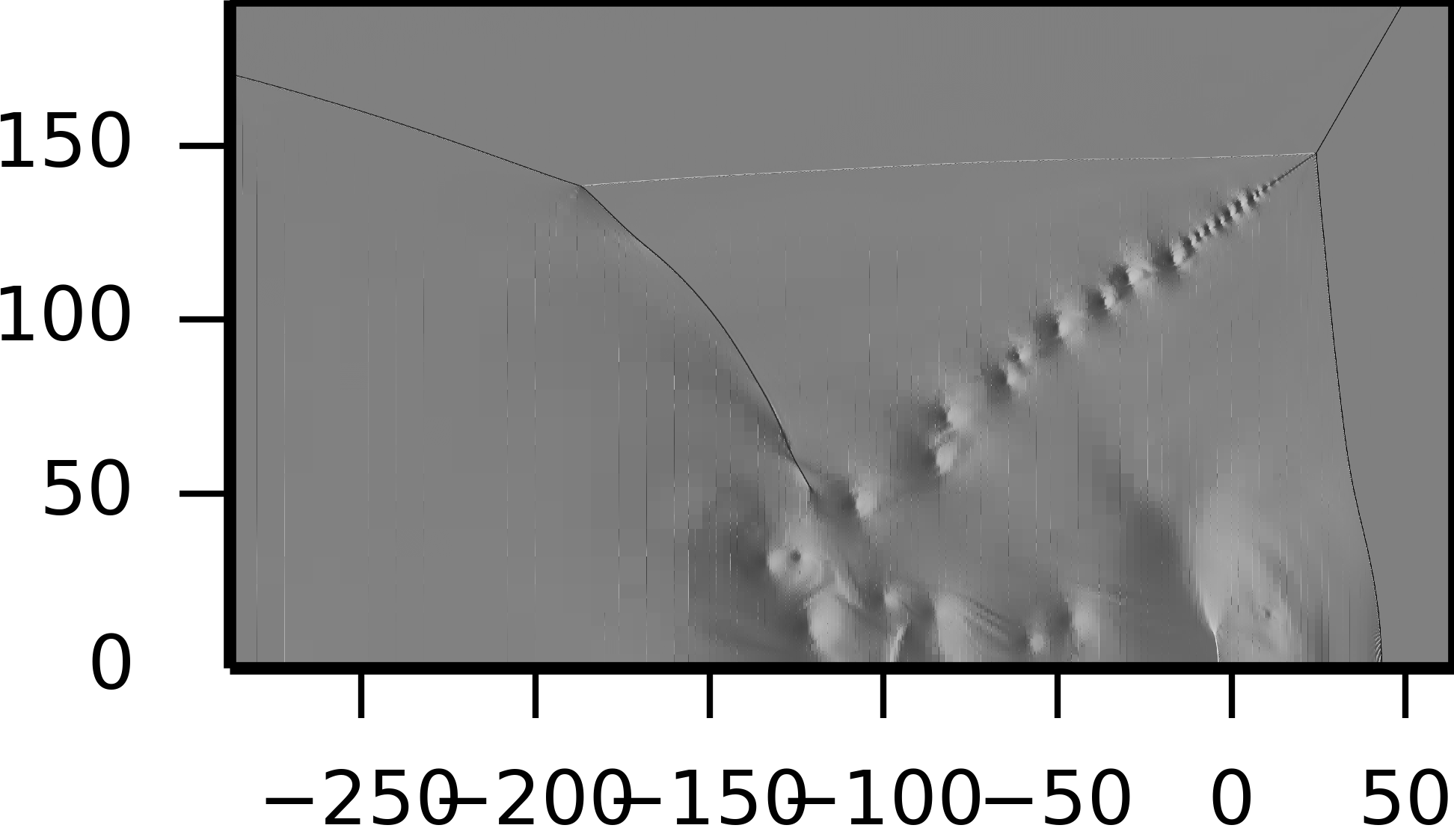}
		\caption{$M_{\mathrm{i}} = 2.5$, $t=363$}
	\end{subfigure}%
	\begin{subfigure}[t]{0.33\textwidth}
		\centering
		\includegraphics[scale=1.02]{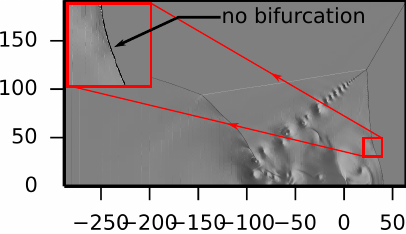}
		\caption{$M_{\mathrm{i}} = 2.75$, $t=265$}
	\end{subfigure}%
	\begin{subfigure}[t]{0.33\textwidth}
		\centering
		\includegraphics[scale=1.02]{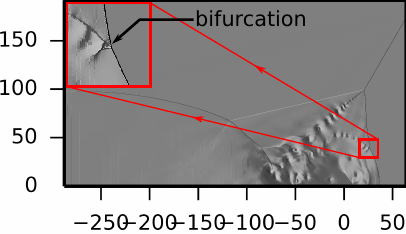}
		\caption{$M_{\mathrm{i}} = 3$, $t=203$}
	\end{subfigure}%
	\caption{\label{fig:bifurcation_resolution_study}Comparison of inviscid simulations across the bifurcation boundary at times equal to viscous triple shock calculations with $Re_{\mathrm{target}} = 100$ (top) and $1000$ (bottom), $\gamma = 1.06$, $\theta_{\mathrm{w}}=30^{\circ}$; insets at the top left of the subfigures are magnifications of the Mach stem}
\end{figure}

A resolution study is first performed, since the lack of scale in the Euler equations make it ambiguous to know how much time the inviscid simulations should be run. 
Figure \ref{fig:bifurcation_resolution_study} shows inviscid simulations near the bifurcation limit at target times calculated using equation \ref{eq:target_time}, with $Re_{\mathrm{target}} = 100$ (top row) and an $Re_{\mathrm{target}} = 1000$ (bottom row). The latter case has about ten times more grid points resolving the Mach stem height.
The schlieren plots are inspected for the presence of a transverse shock wave on the Mach stem, as indicated in figure \ref{fig:bifurcation_resolution_study}, and the Mach stem is considered to have bifurcated if one is found. This is similar to how transitional and double Mach reflections were differentiated in past work. 

No bifurcation is found at $M_{\mathrm{i}} = 2.5$, and a bifurcation is found at $M_{\mathrm{i}} = 3$ at both resolutions. The $M_{\mathrm{i}} = 2.75$ case is critical, with the bifurcation at $t = 26.5$ disappearing by $t = 265$. 
This critical behaviour is to be expected near the bifurcation limit since the phenomenon is sensitive to diffusion and affected by the appearance of Kelvin-Helmholtz instability. Using $Re_{\mathrm{target}} = 100$ to calculate simulation time is sufficient to recover the bifurcation limit found in viscous reflections at $Re \approx 1000$, within $\pm 0.25$ of the Mach number in this case.

\begin{figure}
	\centering
	$\blacksquare$ bifurcation observed\ \ \ \ $\square$ no bifurcation observed\\
	\begin{subfigure}[b]{0.45\textwidth}
		\centering
		\includegraphics[scale=1]{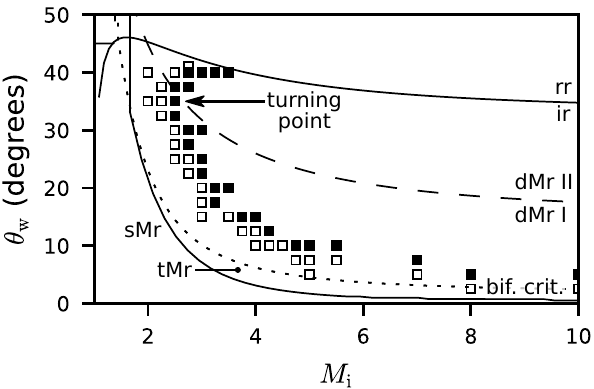}
		\caption{\label{fig:simulation:bifurcation_map:g=1.06}$\gamma=1.06$}
	\end{subfigure}%
	\begin{subfigure}[b]{0.45\textwidth}
		\centering
		\includegraphics[scale=1]{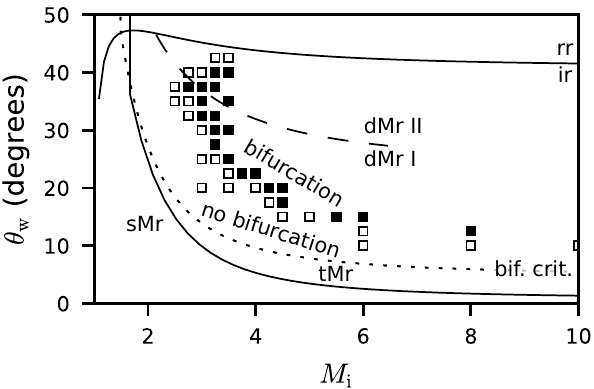}
		\caption{\label{fig:simulation:bifurcation_map:g=1.15}$\gamma=1.15$}
	\end{subfigure}%
	\caption{\label{fig:simulation:bifurcation_map}Mach stem bifurcation domain from inviscid simulations; rr: regular reflection; ir: irregular reflection; sMr: single Mach reflection; tMr: transitional Mach reflection; dMr: double Mach reflection, cases I and II; bif. crit.: analytic bifurcation criterion (\cite{mach_bifurcating_2011}, bifurcations predicted above, no bifurcations below)}
\end{figure}

\begin{figure}
	\centering
	$\blacksquare$ bifurcation observed\ \ \ 
	$\square$ no bifurcation observed\\
	\includegraphics[scale=1]{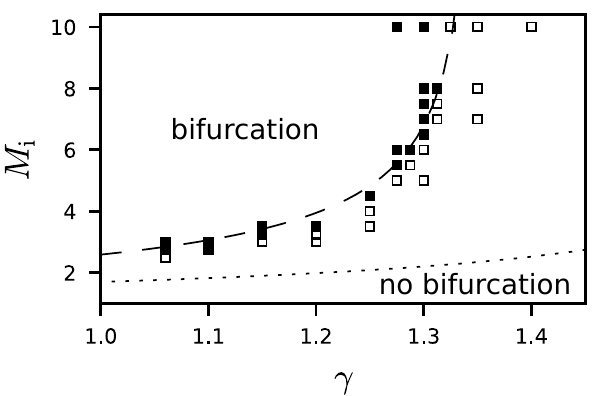}
	\caption{\label{fig:simulation:bifurcation_onset:gamma}Mach stem bifurcation domain from inviscid simulations at a fixed angle ($\theta_{\mathrm{w}}=30^{\circ}$); dotted: analytic bifurcation criterion (\cite{mach_bifurcating_2011}); dashed: shock Mach number for a density ratio of 6.7}
\end{figure}

The computations are extended to a larger range of $M$, $\theta_{\mathrm{w}}$ and $\gamma$ and plotted in figures \ref{fig:simulation:bifurcation_map} and \ref{fig:simulation:bifurcation_onset:gamma}. Each square represents one simulation; filled squares show simulations with a bifurcated Mach stem and hollow squares indicate simulations without one.
The Mach stem bifurcation limit lies on the boundary between solid and hollow squares. 

Figure \ref{fig:simulation:bifurcation_map} shows the effect of Mach number and reflection angle on Mach stem bifurcations. 
The inviscid results match the bifurcation limit of viscous simulations ($2.5 < M_{\mathrm{i}} \le 3$, $\gamma = 1.06$, $\theta_{\mathrm{w}} = 30^{\circ}$) and are in agreement with the onset of Mach stem deformation found in the experiments ($2.9 < M_{\mathrm{c}} \le 3.5$ for $\gamma_0 = 1.15$ and $2.7 < M_{\mathrm{c}} \le 3.4$ for $\gamma_0 = 1.06$, $\theta_{\mathrm{w}} = 30^{\circ}$).
The $\gamma=1.15$ limit also lies close to the experimental limit of $3.84 < M_{\mathrm{i}} \le 4.09$ at $\gamma_0=1.13$ and $\theta_{\mathrm{w}}=15^{\circ}$ found by \cite{semenov_classification_2009b}. 

A number of curves are plotted along with the simulation results in figure \ref{fig:simulation:bifurcation_map} to help situate where Mach stem bifurcations begin relative to other types of shock reflection. The curves delimit other shock reflection boundaries: the sonic criterion of the regular reflection boundary (ir/rr) is plotted at the top with a solid nearly-horizontal curve. Mach reflections do not occur above this curve. 
The lower solid curve connected to the vertical line represents the single/transitional Mach reflection boundary (sMr/tMr). Transitional and double Mach reflections occur above this curve. Details on these limits are available from \cite{ben-dor_shock_2007}.

The dashed curve identifies the boundary between cases I and II of the double Mach reflection (dMr I/II), defined by \cite{li_reconsideration_1995}. Case I lies below the curve, and case II is above.
The flow around the secondary triple point in a double Mach reflection can be calculated using three shock theory, with an assumption about the angles between the shocks.
In a type I double Mach reflection, the secondary reflected shock is assumed to be straight and intersect the slip line at a right angle (\textit{e.g.} figure \ref{fig:simulations:gamma=1.06,M=2.5,Re=1000}). A type II double Mach reflection occurs when this assumption would put the intersection point below the reflecting surface, so the secondary reflected shock is assumed to form a straight line between the secondary triple point and the intersection of the slip line with the reflecting surface.

The dMr I/II boundary fits well with a turning point of the bifurcation limit. \cite{mach_bifurcating_2011} postulated the secondary reflected shock of a type II double Mach reflection would prevent bifurcation, by turning the jet away from the Mach stem, however the transition from I to II only shifts the bifurcation limit to higher Mach numbers. 

The dotted line in figure \ref{fig:simulation:bifurcation_map} (bif. crit.) is an analytic bifurcation criterion proposed by \cite{mach_bifurcating_2011}. It is based on the idea that the slip line must eventually become parallel to the wall (\cite{hornung_regular_1986}). When pressure from the flow deflection process (flow passing the over wedge tip) exceeds pressure from the shock reflection process, a stagnation point is created near the slip line, driving streamlines to its right into the forward jet. The speed of flow in the jet is compared to the Mach stem speed and if the jet is faster than the Mach stem, bifurcation is assumed to occur. \cite{li_analysis_1999} and \cite{shi_mach_2019} performed similar analyses to account for bulging (not bifurcation) of the Mach shock. 
This criterion underestimates the Mach number and wedge angle required for bifurcation and performs poorly at higher isentropic exponents.

The dependence of the bifurcation boundary on the isentropic exponent is explored in $M_{\mathrm{i}}$-$\gamma$ phase space in figure \ref{fig:simulation:bifurcation_onset:gamma} for $\theta_{\mathrm{w}} = 30^{\circ}$. 
The Mach number required for bifurcation increases with isentropic exponent until $\gamma \approx 1.3$ is reached. Strong jets that bulge the Mach stem are still present above this value, but no bifurcations are observed despite Mach's bifurcation criterion (dotted curve) predicting otherwise. 
This lack of limit in Mach's bifurcation criterion suggests there is a missing link between jetting and bifurcation. A better criterion is required for predicting the triple Mach-White reflections.

\subsection{\label{sec:Implication to detonation structure}Implication to detonation structure}

Shock reflections have been shown to contribute to the propagation of detonations (\cite{lee_detonation_2008}). Detonations are supersonic combustion waves that suffer from a multi-dimensional instability that causes their shock front to be punctuated by triple points. The triple points periodically reflect from each other, creating new triple points. A curved Mach stem is formed upon reflection, like in the experiments, and the new triple points go on to collide with their neighbours. The Mach stem becomes the incident shock for the next reflection and the process repeats. The experiments' curved shocks form a nice analogue to the unsteady shock reflection process in detonations.

Tracing the triple point paths of a detonation creates a pattern of interlocking diamonds called the cellular detonation structure. Detonations are often characterized by the size and regularity of their cellular structure.
Detonations with irregular structures are more resistant to quenching and their mixtures are easier to detonate than regular ones (\cite{moen_influence_1986,radulescu_failure_2002}). 
\cite{radulescu_origin_2009,mach_mach_2011,mach_bifurcating_2011} suggested that Mach stem bifurcation may contribute to the irregularity of the detonation cellular structure by creating new detonation cells. They showed that bifurcations contribute to an irregular appearance on numerical soot foils of detonations. They tabulated 21 experiments from other work and categorized the cellular structure as regular, intermediate, and irregular and found perfect agreement between regular detonations and the absence of bifurcation in their inviscid simulations. A clear distinction can be made from their tabulated data: irregular and intermediate structures arise when $\gamma_1 \leq 1.32$ and regular structures occur when $1.41 \leq \gamma_1$. This closely matches the bifurcation limit of $\gamma \approx 1.3$ found in this study. 

The detonability and quenching resistance exhibited by irregular mixtures may be a result of the strong jetting and large-scale mixing that occurs behind the Mach stem in gases with low isentropic exponents. 
Experiments show this mixing becomes especially evident when Kelvin-Helmholtz instabilities are entrained into the vortex. In detonations, reaction rates behind the Mach stem will be sensitive to the presence of mixing and turbulence, predominantly near detonation propagation limits.
A parametric investigation of the effect of isentropic exponent on mixing in detonations and their cellular structure would be of particular interest.

\subsection{\label{sec:Shock instability at high Mach number and low isentropic exponent}Shock instability at high Mach number and low isentropic exponent}

The textured appearance of the schlierens behind the incident, reflected, and Mach shocks seen in certain experiments (\textit{e.g.} figures \ref{fig:experiment:2017-04-06-C:14}, \ref{fig:experiment:2017-03-30-D:17}, and \ref{fig:experiment:2017-03-30-G:15}) indicates a non-uniform density field. The flow field becomes less uniform as the isentropic exponent is decreased from $\gamma_0 = 1.15$ to 1.06 and as the Mach number is increased. 
The non-uniformity is not caused by contamination of the test section by driver gas. This is evidenced by the fact the bow shocks remain intact and attached to the chevron, as discussed in section \ref{sec:Experiments at higher Mach number}. Furthermore, simulations in the appendix show that the driver/test gas interface is far from the leading shock in all experiments when the full stand-off distance of $\hat{l}_{\mathrm{d}} = 1.85$ is used.
The cause of the phenomenon is not understood but may be linked to shock wave instability. 

Shock wave instability has been ascribed to endothermic processes such as ionization (\cite{grun_instability_1991,glass_effects_1978,griffiths_stability_1976}), dissociation (\cite{griffiths_stability_1976}), or to heavy gases (\textit{i.e.} low isentropic exponent gases) where vibrational relaxation may occur behind the shock (\cite{griffiths_stability_1976,mishin_anomalous_1981,hornung_shock_2001,semenov_classification_2009b,ohnishi_bow-shock_2015}). Shock instability has been observed by \cite{sirmas2015evolution,sirmas2018structure} in molecular dynamic calculations of relaxing shock waves in gases with inelastic collisions, as would occur where strong vibrational relaxation effects are present.
Work with hypersonic projectiles have attributed similar shock instability to vibrating or unstable contact surfaces. 
Experiments and simulations of \cite{hornung_shock_2001} and \cite{ohnishi_bow-shock_2015} found instability when density jumps of ${\rho_{\mathrm{1}}}/{\rho_0} \ge 14$ and ${\rho_{\mathrm{1}}}/{\rho_0} \ge 10$ were reached across the shock. 
Interestingly, the bifurcation boundary in $M$-$\gamma$ space is well represented by a constant density ratio of ${\rho_{\mathrm{1}}}/{\rho_0} = 6.7$ across the shock, plotted with a dashed curve in figure \ref{fig:simulation:bifurcation_onset:gamma}.

The stability of these low isentropic exponent gases is of interest and should be investigated in future work. This could be done by introducing vibrational relaxation effects into simulations, for instance, as the instability was not seen in the Navier-Stokes simulations.

\section{Conclusion}\label{sec:conclusion}

Experiments and simulations were used to study the effect of the isentropic exponent, Mach number and Reynolds number on the large-scale convective mixing caused by the forward jet, vortex, and flow behind the Mach stem. Experiments were performed over a limited range of Mach numbers at large Reynolds numbers ($Re \sim 10^5$) and various isentropic exponents. Viscous simulations were performed for a larger range of Mach numbers but were limited to small Reynolds numbers ($Re \sim 10^3$) and $\gamma = 1.06$. Inviscid simulations were used to bridge the gap between viscous simulations and experiments, however, differences between simulations and experiments remain to be reconciled.

In all cases, as the isentropic exponent decreased, or the Mach number or Reynolds number increased, the forward jet approached the Mach stem and developed a vortex. The jet and vortex can grow strong enough to completely disrupt the flow field behind the Mach shock, and cause the Mach stem to bulge once jet-shock interaction becomes important.

The region behind the Mach stem continuously evolves as Reynolds number is increased. A turbulent structure becomes visible behind the Mach stem in experiments when the isentropic exponent is sufficiently small and the Mach number is sufficiently large. Similar heterogeneous flow fields are observed in inviscid simulations. This is attributed to Kelvin-Helmholtz instability in the vortex, and is absent from the viscous results. The experiments clearly show that large-scale convective mixing behind the Mach stem is driven primarily by low isentropic exponents.  When the isentropic exponent is too large, \textit{e.g.} $\gamma=1.4$, neither jet-shock interaction nor large-scale mixing are observed.

The limits of Mach stem bifurcation (triple Mach-White reflection), have been reported in $\theta_{\mathrm{w}}$-$M_{\mathrm{i}}$-$\gamma$ phase space. Bifurcations are found to be absent when $\gamma \gtrsim 1.3$ at $\theta_{\mathrm{w}}=30^{\circ}$, a limit which corresponds closely to the boundary between regular and irregular cellular structures of detonations. 
The detonability and quenching resistance of irregular mixtures may be a result of the strong jetting and large-scale mixing that occurs behind the Mach stem in gases with low isentropic exponents.

\section*{}

The authors thank Sam Falle from the University of Leeds for generously allowing the use of his computational code \verb|mg| which made the numerical simulations possible.

The authors acknowledge financial support from the Natural Sciences and Engineering Research Council of Canada (NSERC) through the Discovery Grant ``Predictability of detonation wave dynamics in gases: experiment and model development'' (2017-2022).

Declaration of Interests: The authors report no conflict of interest.

\bibliographystyle{jfm}
\bibliography{references}

\appendix 
\vspace{-0.5cm}
\section{\label{sec:numerical demonstration of experiments}Simulations with varying diaphragm stand-off distances}

Inviscid simulations of the shock tube and obstacle were performed to assess the diaphragm stand-off distance and to gauge the effect of the driver/test gas interface on the shock reflection.

The two-dimensional domain spanned half the shock tube height, from the centre-line to the shock tube wall, five half-heights upstream of the chevron tip, and three half-heights downstream. The domain was covered by a $40\times5$ coarse grid with six levels of refinement for a finest possible grid of $2560\times320$ and resolution of 3.15 grid points per millimetre. The chevron was created with a step-like boundary. Step boundaries have been shown to affect the shock reflection configuration (\cite{ben-dor_influence_1987}), but perform decently (\cite{falle_body_1992}) with sufficient dissipation (controlled here by low resolution). The top and bottom boundaries employed symmetry conditions, and the left and right boundary had zero normal gradients. The origin is located at the trailing edge of the chevron.
The simulations were initiated with a shock located at $\hat{x}_{\mathrm{s,0}} = - 15.24$ cm. The driver gas was initially at $\hat{x}_{\mathrm{d,0}} = \frac{u_{\mathrm{t}}}{D_{\mathrm{t}}} (\hat{x}_{\mathrm{s,0}} - \hat{l}_{\mathrm{d}})  + \hat{x}_{\mathrm{s,0}}$, where $D_{\mathrm{t}}=M_{\mathrm{t}} c_0$ is the shock speed and $u_{\mathrm{t}}$ is the shocked particle speed. 
The driver gas has the same pressure and velocity as the shocked gas, but with a density of $2\rho_0$.

Experiment 9 (see table \ref{tab:experimental conditions}, $M_{\mathrm{c}}=3.4$ and $\gamma = 1.06$) is simulated because it is the most compressible (\textit{i.e.} highest Mach number and lowest isentropic exponent) of all experiments with a stand-off of $\hat{l}_{\mathrm{d}} = 1.85$ m. If this experiment is unaffected by the driver/test gas interface, then so are all the other experiments where $1.06 \le \gamma_0$, $M_{\mathrm{c}} \le 3.4$, and $\hat{l}_{\mathrm{d}} = 1.85$ m. An initial shock strength of $M_{\mathrm{t}} = 4.2$ is required to generate a shock of $M_{\mathrm{c}}=3.4$ on the chevron.

The simulation results are presented in figures \ref{fig:simulation:chevron:g=1.06,Mc=3.4,ld=1.85}. The figures show schlieren images on the top halves. The bottom halves show temperature cutoff at $T=1.6$ to emphasize details in the reflection. The $x$ and $y$ coordinates are scaled in metres.

Figure \ref{fig:simulation:chevron::g=1.06,Mc=3.4,ld=1.85:0} shows the initial conditions, \textit{i.e.} after a shock has been transmitted through the diaphragm, but before it interacts with the obstacle.
The driver/test gas interface is overlayed with a dotted red line. 
The shock diffracts over the chevron and reaches the trailing edge in figure \ref{fig:simulation:chevron::g=1.06,Mc=3.4,ld=1.85:51}. 
The driver/test gas interface remains far behind the shock front.
The shock wave reflects from the axis of symmetry, forming a double Mach reflection with Mach stem bifurcation, as shown in figure \ref{fig:simulation:chevron::g=1.06,Mc=3.4,ld=1.85:92}.
The driver gas has diffracted over the obstacle, driving pressure a wave that has not reached the front. 
When there is no driver/test gas interface (figure \ref{fig:simulation:chevron::g=1.06,Mc=3.4,ld=infty:92}), the shock front is identical that of figure \ref{fig:simulation:chevron::g=1.06,Mc=3.4,ld=1.85:92}, showing the driver/test gas interface does not interfere with the reflection front when $\hat{l}_{\mathrm{d}} = 1.85$ m.

Figure \ref{fig:simulation:chevron::g=1.06,Mc=4.0,ld=0.68:92} is a simulation of experiment 11 ($M_{\mathrm{c}}=4.0$ and $\gamma = 1.06$) where the stand-off distance was shortened to $\hat{l}_{\mathrm{d}}=0.68$ m. The driver/test gas interface is closer to the shock front, the double Mach reflection has a Mach stem that is clearly bifurcated and taller than the previous case, despite the higher Mach number. The bow shock (secondary Mach shock) travels much faster through the hot driver gas, forming a vertical shock.

These simulations demonstrate that the driver/test gas interface has no effect on the experiments with $\hat{l}_{\mathrm{d}} = 1.85$ m, and that only experiments 10 and 11, where $\hat{l}_{\mathrm{d}} < 1.85$ m, suffer from interference by the driver gas.

\begin{figure}
	\centering
	\begin{subfigure}[t]{1\textwidth}
		\centering
		\includegraphics[scale=0.6]{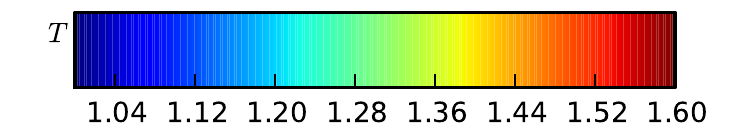}
	\end{subfigure}
	\begin{subfigure}[t]{1\textwidth}
		\centering
		\includegraphics[width=\textwidth]{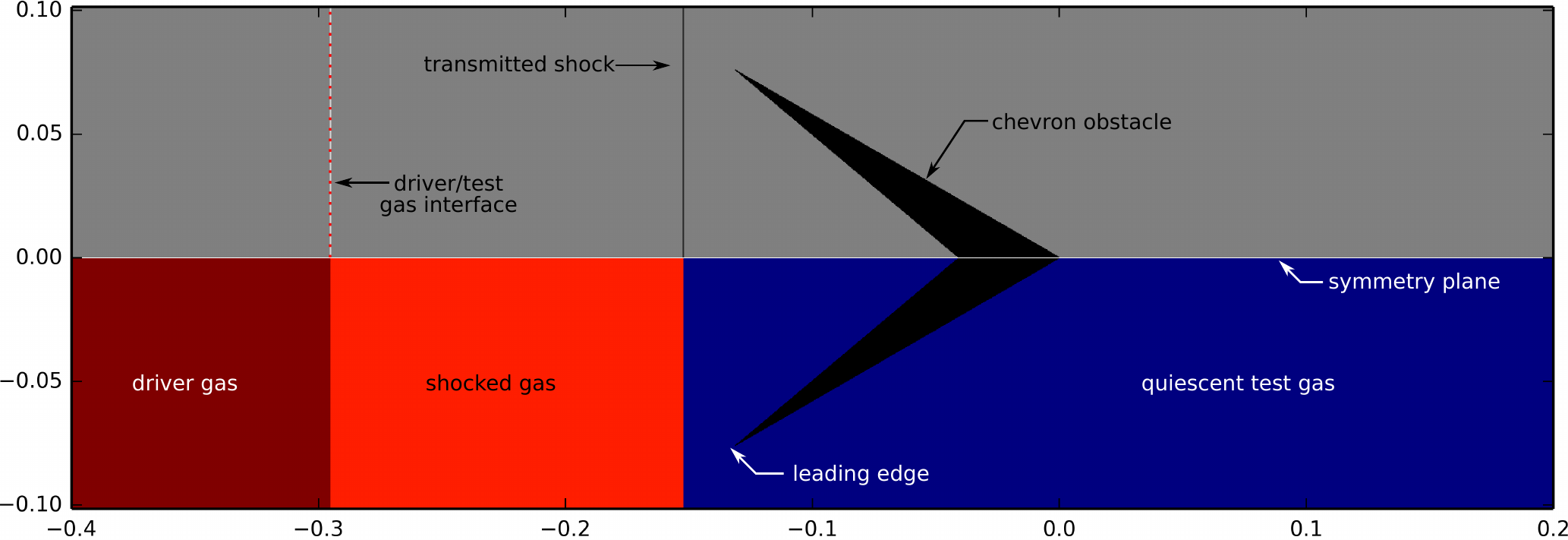}
		\caption{\label{fig:simulation:chevron::g=1.06,Mc=3.4,ld=1.85:0}Initial condition, before shock interaction with chevron}
	\end{subfigure}
	\\
	\begin{subfigure}[t]{0.5\textwidth}
		\centering
		\includegraphics[width=\textwidth]{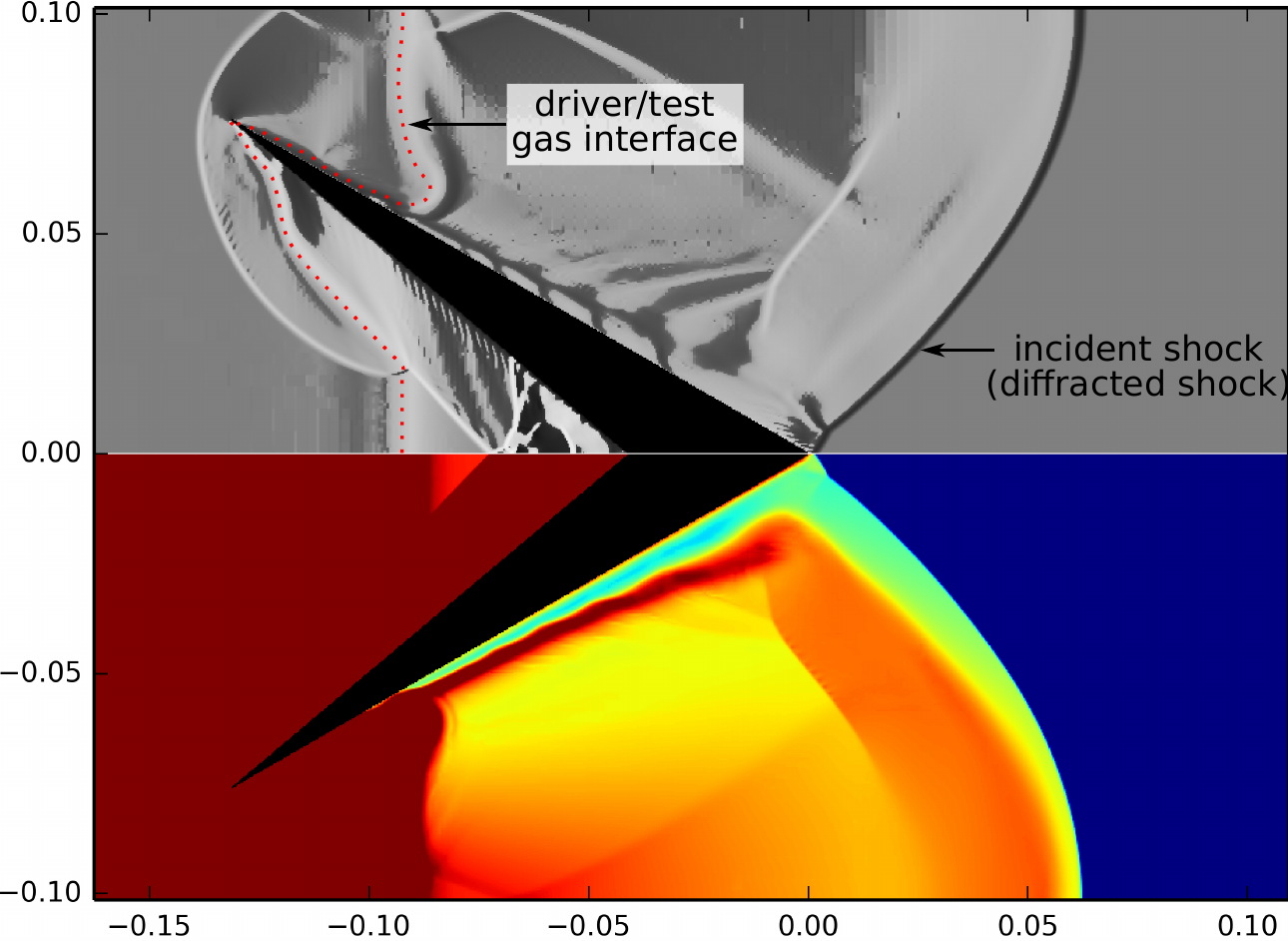}
		\caption{\label{fig:simulation:chevron::g=1.06,Mc=3.4,ld=1.85:51}Shock diffracting over chevron}
	\end{subfigure}%
	\begin{subfigure}[t]{0.5\textwidth}
		\centering
		\includegraphics[width=\textwidth]{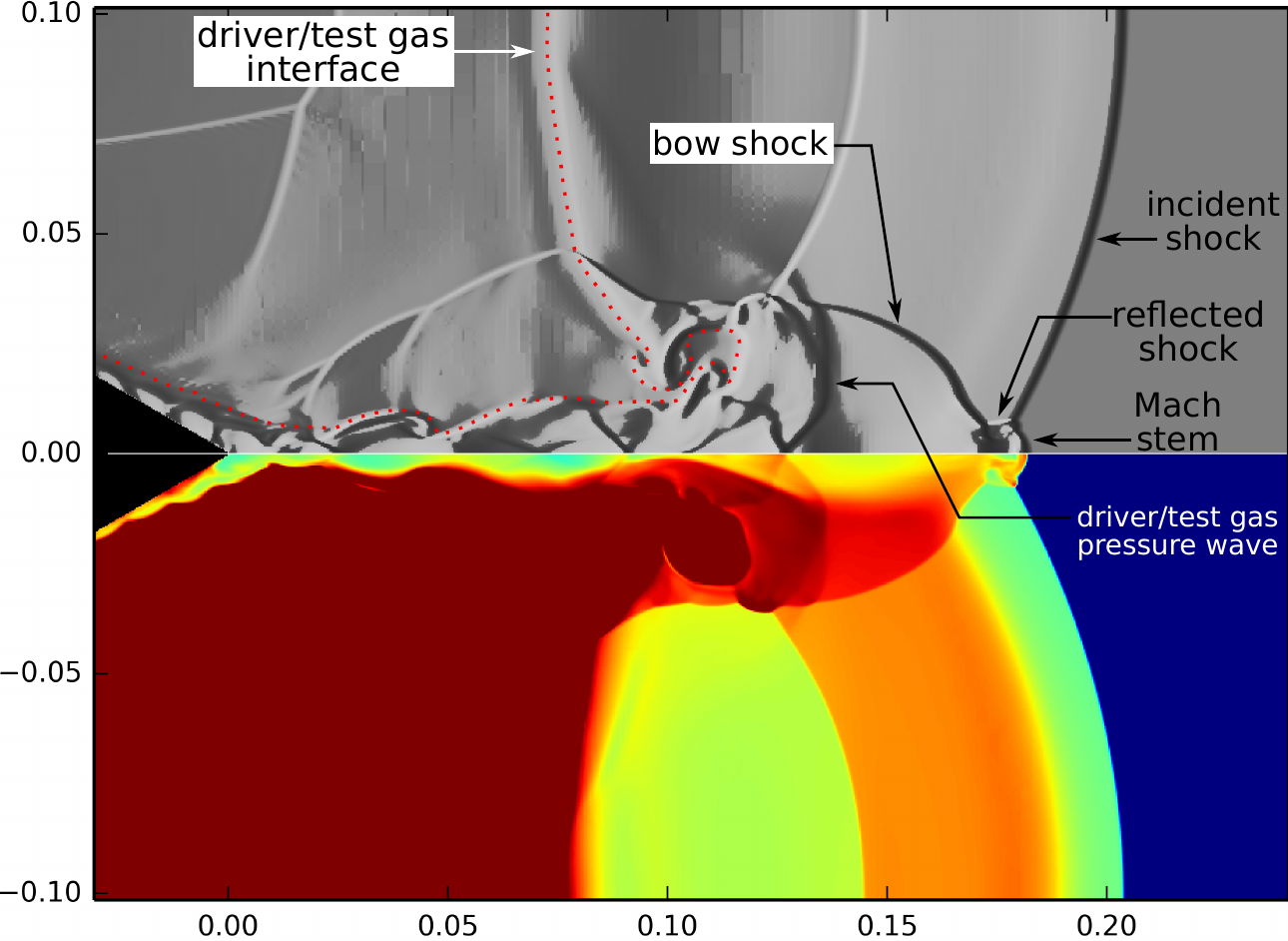}
		\caption{\label{fig:simulation:chevron::g=1.06,Mc=3.4,ld=1.85:92}Shock reflection from plane of symmetry}
	\end{subfigure}
	\caption{\label{fig:simulation:chevron:g=1.06,Mc=3.4,ld=1.85}Inviscid simulation of shock reflection over a chevron ($\gamma = 1.06$, $M_{\mathrm{c}}=3.4$, $\hat{l}_{\mathrm{d}}=1.85$ m); top: schlieren; dotted red line: driver/test gas interface; bottom: temperature; axes scaled in metres}
\end{figure}

\begin{figure}
	\centering
	\begin{subfigure}[t]{0.5\textwidth}
		\centering
		\includegraphics[width=\textwidth]{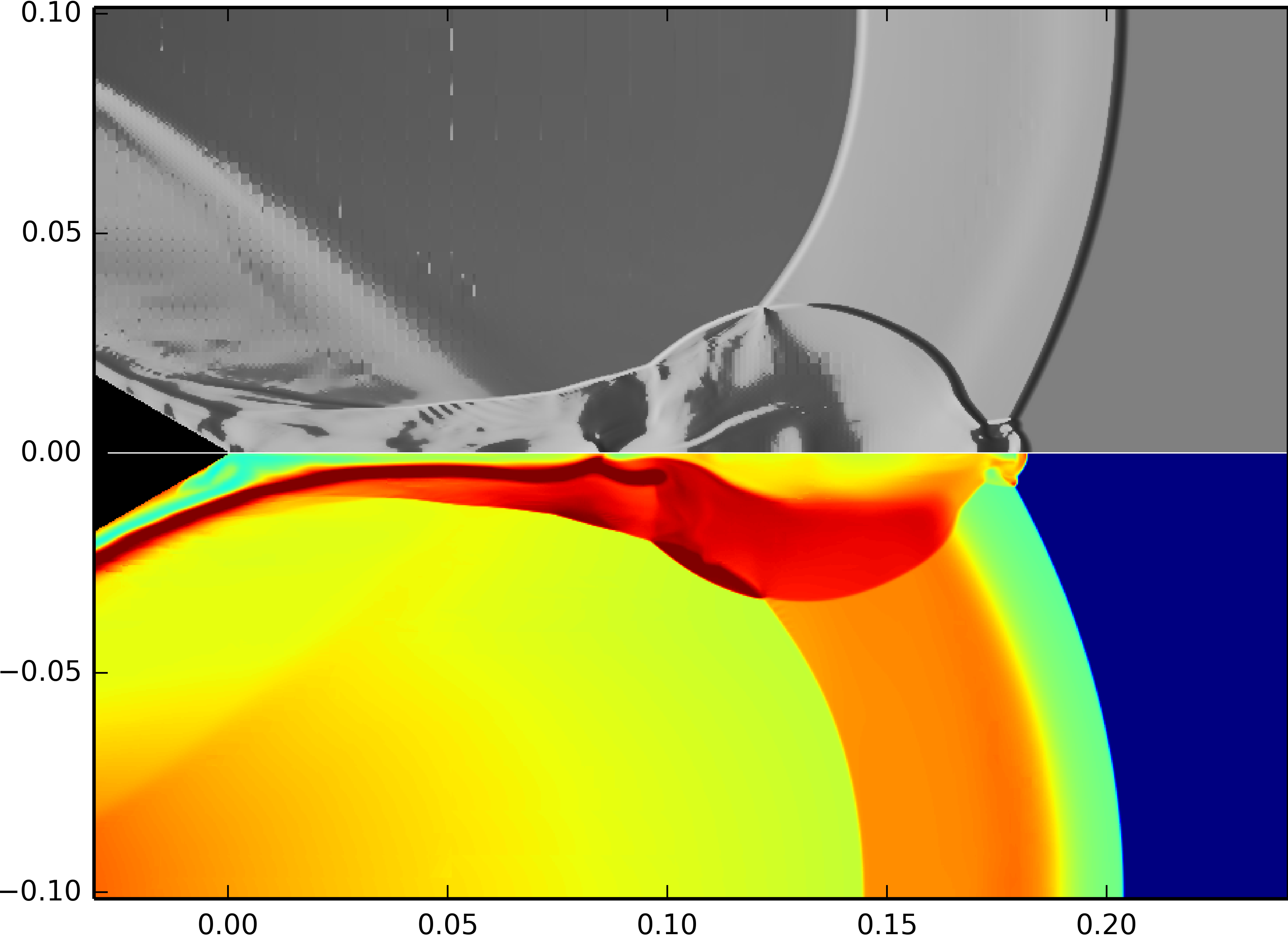}
		\caption{\label{fig:simulation:chevron::g=1.06,Mc=3.4,ld=infty:92}$M_{\mathrm{c}}=3.4$, no driver/test gas interface}
	\end{subfigure}%
	\begin{subfigure}[t]{0.5\textwidth}
		\centering
		\includegraphics[width=\textwidth]{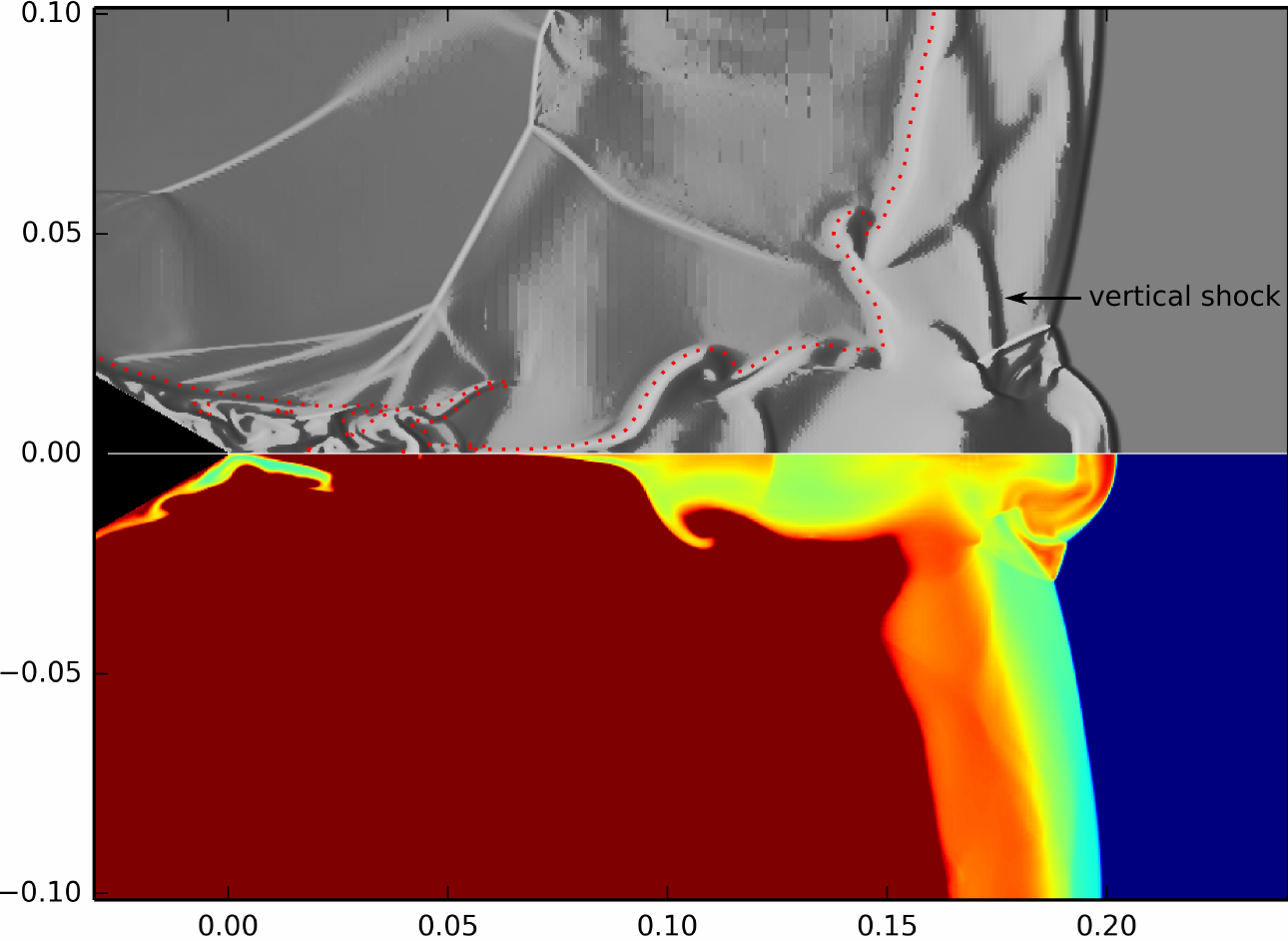}
		\caption{\label{fig:simulation:chevron::g=1.06,Mc=4.0,ld=0.68:92}$M_{\mathrm{c}}=4.0$, $\hat{l}_{\mathrm{d}}=0.68$ m}
	\end{subfigure}%
	\caption{\label{fig:simulation:chevron:other}Inviscid simulations of shock reflection over a chevron ($\gamma = 1.06$); top: schlieren; dotted red line: driver/test gas interface; bottom: temperature; axes scaled in metres}
\end{figure}

\end{document}